\documentclass[twocolumn]{aastex631}
\usepackage{natbib,amsmath}
\usepackage{xspace}
\usepackage{subfigure}

\providecommand{\e}[1]{\ensuremath{\times 10^{#1}}}

\begin{document}

\title{Constraining atmospheric composition from the outflow: helium observations reveal the fundamental properties of two planets straddling the radius gap}

\correspondingauthor{Michael Zhang}
\email{mzzhang2014@gmail.com}

\author[0000-0002-0659-1783]{Michael Zhang}
\altaffiliation{51 Pegasi b Fellow}
\affil{Department of Astronomy \& Astrophysics, University of Chicago, Chicago, IL 60637}

\author[0000-0003-4733-6532]{Jacob L.\ Bean}
\affil{Department of Astronomy \& Astrophysics, University of Chicago, Chicago, IL 60637}

\author[0000-0001-9667-9449]{David Wilson}
\affiliation{Laboratory for Atmospheric and Space Physics, University of Colorado Boulder}

\author[0000-0002-7119-2543]{Girish Duvvuri}
\affiliation{Vanderbilt University}

\author[0000-0002-5094-2245]{Christian Schneider}
\affiliation{Hamburger Sternwarte, Universität Hamburg, Gojenbergsweg 112, D-21029 Hamburg, Germany}

\author[0000-0002-5375-4725]{Heather A. Knutson}
\affiliation{Division of Geological and Planetary Sciences, California Institute of Technology}

\author[0000-0002-8958-0683]{Fei Dai}
\affiliation{Institute for Astronomy, University of Hawai`i, 2680 Woodlawn Drive, Honolulu, HI 96822, USA}

\author[0000-0001-6588-9574]{Karen A.\ Collins}
\affiliation{Center for Astrophysics \textbar \ Harvard \& Smithsonian, 60 Garden Street, Cambridge, MA 02138, USA}

\author[0000-0001-8621-6731]{Cristilyn N.\ Watkins}
\affiliation{Center for Astrophysics \textbar \ Harvard \& Smithsonian, 60 Garden Street, Cambridge, MA 02138, USA}

\author[0000-0001-8227-1020]{Richard P. Schwarz}
\affiliation{Center for Astrophysics \textbar \ Harvard \& Smithsonian, 60 Garden Street, Cambridge, MA 02138, USA}

\author[0000-0003-1464-9276]{Khalid Barkaoui}
\affiliation{Astrobiology Research Unit, Universit\'e de Li\`ege, 19C All\'ee du 6 Ao\^ut, 4000 Li\`ege, Belgium}
\affiliation{Department of Earth, Atmospheric and Planetary Science, Massachusetts Institute of Technology, 77 Massachusetts Avenue, Cambridge, MA 02139, USA}
\affiliation{Instituto de Astrof\'isica de Canarias (IAC), Calle V\'ia L\'actea s/n, 38200, La Laguna, Tenerife, Spain}

\author[0000-0002-1836-3120]{Avi Shporer}
\affiliation{Department of Physics and Kavli Institute for Astrophysics and Space Research, Massachusetts Institute of Technology, Cambridge, MA 02139, USA}

\author[0000-0003-1728-0304]{Keith Horne}
\affiliation{SUPA School of Physics and Astronomy, University of St\,Andrews, St\,Andrews, Fife, KY16\,9SS Scotland, UK}

\author[0000-0003-3904-6754]{Ramotholo Sefako} 
\affiliation{South African Astronomical Observatory, P.O. Box 9, Observatory, Cape Town 7935, South Africa}

\author[0000-0001-9087-1245]{Felipe Murgas}
\affiliation{Instituto de Astrof\'isica de Canarias (IAC), E-38205 La Laguna, Tenerife, Spain}
\affiliation{Departamento de Astrof\'isica, Universidad de La Laguna (ULL), E-38206 La Laguna, Tenerife, Spain}

\author[0000-0003-0987-1593]{Enric Palle}
\affiliation{Instituto de Astrof\'\i sica de Canarias (IAC), 38205 La Laguna, Tenerife, Spain}
\affiliation{Departamento de Astrof\'\i sica, Universidad de La Laguna (ULL), 38206, La Laguna, Tenerife, Spain}

\begin{abstract}
TOI-836 is a $\sim2-3$ Gyr K dwarf with an inner super Earth ($R=1.7\,R_\Earth$, $P=3.8$\,d) and an outer mini Neptune ($R=2.6\,R_\Earth$, $P=8.6$\,d).  JWST/NIRSpec 2.8--5.2 $\mu$m transmission spectra are flat for both planets.  We present Keck/NIRSPEC observations of escaping helium for super-Earth b, which shows no excess absorption in the 1083 nm triplet to deep limits ($<0.2$\%), and mini-Neptune c, which shows strong (0.7\%) excess absorption in both visits.  These results demonstrate that planet c retains at least some primordial atmosphere, while planet b is consistent with having lost its entire primordial envelope.  Self-consistent 1D radiative-hydrodynamic models of planet c reveal that the helium excess absorption signal is highly sensitive to metallicity: its equivalent width collapses by a factor of 13 as metallicity increases from 10x to 100x solar, and by a further factor of 12 as it increases to 200x solar.  The observed equivalent width is 88\% the model prediction for 100x metallicity, suggesting an atmospheric metallicity similar to K2-18b and TOI-270d, the first two mini-Neptunes with detected absorption features in JWST transmission spectra.  We highlight the helium triplet as a potentially powerful probe of atmospheric composition, with complementary strengths and weaknesses to atmospheric retrievals. The main strength is its extreme sensitivity to metallicity in the scientifically significant range of 10--200x solar, and the main weakness is the enormous model uncertainties in outflow suppression and confinement mechanisms, such as magnetic fields and stellar winds, which can suppress the signal by at least a factor of $\sim$several.
\end{abstract}


\section{Introduction} 
\label{sec:intro}
The planets between 1 and 3.5 $R_\Earth$ are possibly the most common class of planet in existence \citep{zhu_2021}.  None are present in the solar system, making their properties fundamentally mysterious.  The radius gap at 1.7 $R_\Earth$ \citep{fulton_2017,fulton_2018} that divides super-Earths from mini-Neptunes has prompted copious research on its origins and on the fundamental properties of planets below, inside, and above the gap (c.f. \citealt{bean_2021,wordsworth_2022}).  It is evident from their densities that super-Earths cannot have a thick primordial hydrogen/helium envelope, but thin or high mean molecular weight atmospheres cannot be ruled out based on mass and radius alone.  The lower densities of mini-Neptunes and of some planets inside the radius gap can be explained by thick primordial envelopes \citep{rogers_2023}, large amounts of water (e.g. \citealt{mousis_2020}), a mixture of both types of planet \citep{luque_2022}, or a mixture of both types of gas in the same planets (e.g. \citealt{benneke_2024}).

If the radius gap is indeed due to the prevalence or absence of a primordial envelope, suggested explanations of how super-Earths have lost their envelopes include photoevaporation \citep{owen_2017,mills_2017} and core-powered mass loss \citep{ginzburg_2018,gupta_2019}.  Recent work combining the two mechanisms suggests that both are important in different parts of parameter space, but that photoevaporation is responsible for the final carving of the radius valley \citep{owen_2024}.  One major caveat with these population-level studies is that they assume solar metallicity envelopes.  As we shall see, if the envelopes actually have 100-200x solar metallicity, the picture could change because mass loss rates decrease dramatically.

Observations of mass loss from mini-Neptunes have shed light on their composition and evolution.  In 2018, theoretical work by \cite{oklopcic_2018} and observations by \cite{spake_2018} opened up the metastable helium triplet at 1083 nm as an observational probe of escaping atmospheres.  Since then, dozens of exoplanets have been detected in helium absorption, of which six are mini-Neptunes.  In our survey of young mini Neptunes, we saw helium absorption in all of our first four targets \citep[TOI-560b, -1430b, -1683b, and -2076b;][]{zhang_2023}.  The absorption from TOI-1430b was subsequently confirmed by \cite{orell-miquel_2023}, and two mature mini-Neptunes have also been found to have escaping helium: TOI-2134b \citep{zhang_2023b} and -2018b \citep{orell-miquel_2024}.  The ubiquity of outflows shows that most mini Neptunes orbiting Sun-like G and K stars retain at least some of their primordial atmospheres, while the width of the helium absorption suggests that the outflows are due to photoevaporation, not core-powered mass loss \citep{zhang_2023}.

In parallel to the mass loss research, the James Webb Space Telescope (JWST) has enabled far more precise spectra of mini-Neptunes.  The first JWST exoplanet phase curve was of GJ\,1214b, and comparison of the phase curve with general circulation models suggests an atmospheric metallicity equal to or greater than 100x solar \citep{kempton_2023,gao_2023}.  Similarly high metallicities are inferred from the NIRISS+NIRSpec transmission spectrum of K2-18b ($\sim100\times$ solar; \citealt{madhusudhan_2023,wogan_2024}) and TOI-270d ($225_{-86}^{+98}\times$, \citealt{benneke_2024}).  \cite{benneke_2024} suggests that while some mini-Neptunes are Hycean worlds with stratified layers of rock, ice, liquid water, and hydrogen/helium gas, the vast majority of known mini-Neptunes have miscible envelopes that (except for the topmost portion) are in a supercritical state.  They find that even without ice accretion during formation, these envelopes would be highly enriched in water due to chemical reactions between the hydrogen in the envelope and the magma ocean.  JWST continues to observe mini-Neptunes to determine their fundamental atmospheric properties, but not every planet reveals its secrets as easily.  For example, the NIRSpec/G395H transmission spectrum of the warm TOI-836c is flat \citep{wallack_2024}, as is the spectrum of its inner super-Earth companion, TOI-836b \citep{alderson_2024}.  The emerging pattern that warmer mini-Neptunes are featureless while colder mini-Neptunes are not, if not due to bad luck and small number statistics, could indicate that the former have thicker hazes--hazes that would make their characterization difficult.

In this paper, we characterize the basic properties of the TOI-836 system (Section\,\ref{sec:system_properties}), then use observations of escaping helium (Sections \ref{sec:observations} and \ref{sec:results}) and simulations (Section \ref{sec:modelling}) to characterize and compare the atmospheres of TOI-836b and c (Section\,\ref{sec:discussion_conclusion}).  Detecting helium means a planet has retained at least some primordial atmosphere, and cannot be entirely a water world; not detecting it to deep limits suggests it has no primordial atmosphere.  We use an improved version of the self-consistent 1D radiation hydrodynamics code The PLUTO-CLOUDY Interface (TPCI; \citealt{salz_2015}) to show that the helium signal falls sharply with metallicity between 10x and 200x solar, and propose the equivalent width of the helium line as an alternative probe of atmospheric metallicity (Section\,\ref{sec:discussion_conclusion}).  The trend of high metallicity suppressing outflows has been previously noticed both by us \citep{zhang_2022b} and \cite{linssen_2024}, but neither work explored metallicities as high as 200x solar--metallicities which JWST data indicate may be common among mini-Neptunes.  We conclude by discussing the strengths and limitations of this alternative metallicity probe.  

\section{Stellar and planetary properties}
\label{sec:system_properties}
TOI-836 is a bright ($J$=7.6 mag) K dwarf with two known transiting exoplanets, both discovered by TESS \citep{hawthorn_2023}.  The inner planet is a super-Earth with a radius of $1.70 \pm 0.07 R_\Earth$ and mass of $4.5 \pm 0.9 M_\Earth$, while the outer planet is a mini-Neptune with a radius of $2.59 \pm 0.09 R_\Earth$ and mass of $9.6 \pm 2.6 M_\Earth$.  The densities of the two planets are typical of their respective classes.  Interior modelling by \cite{hawthorn_2023} suggests that the inner planet has a rocky composition, while the outer planet requires substantial water or gas to explain its low density.  Interestingly, planet b is almost exactly in the middle of the Fulton radius gap \citep{fulton_2018} (although its high density suggests that it belongs more naturally with the super-Earths), while c is squarely in the mini-Neptune peak.

\subsection{Age}
We observed TOI-836c with Keck because it was initially part of our survey of escaping helium from young mini-Neptunes.  We believed it was young because the two sectors of TESS light curves available for the star both show $\sim$1\% peak-to-trough variability, seemingly with a period of 12 days, which would suggest an age of a few hundred Myr.  Well after we selected the target, the discovery paper \citep{hawthorn_2023} was published, which used two different evolutionary models to derive a combined age estimate of $5.4_{-5.0}^{+6.3}$ Gyr.  This is consistent with every age except the very young.  

\cite{hawthorn_2023} also use 8 years of WASP-South photometry to show that the real rotation period of the star is $22 \pm 0.1$ d, which would be indicative of a mature star.  Phase-folded WASP-South light curves show a peak-to-trough variability of $\sim$1\%, consistent with TESS observations.  From the rotation period and effective temperature, we use gyro-interp \citep{bouma_2023} to estimate an age of $2560_{-130}^{+200}$ Myr.  This estimate should be treated with caution because gyro-interp works by interpolating open cluster data, but only two clusters older than 1 Gyr are in the database: the 2.5 Gyr NGC-6819 and the 2.7 Gyr Rup-147.  These clusters, even when combined, have very few late K dwarfs with measured rotation periods.  Those handful of late K dwarfs have a rotation rate of $\sim$20 d, suggesting (insofar as they are representative of their spectral type) that TOI-836 is of a similar age.

A 17 ks XMM-Newton observation of TOI-836 was taken by a joint HST-XMM program (HST GO 16701, PI: A. Youngblood).  The 0.2--2.5 keV X-ray flux it registered, $8 \times 10^{-14}$ erg\,cm$^{-2}$\,s$^{-1}$ (see Subsection \ref{subsec:high_energy_spec}), implies $L_X=9 \times 10^{27}$ erg\,s$^{-1}$ and $R_X=L_X/L_{\rm bol}=1.6 \times 10^{-5}$.  This ratio is far below the saturation value of $\sim6 \times 10^{-4}$, but an order of magnitude above that of quiet stars like the Sun \citep{wright_2011,jackson_2012}.  Compared to the other five hosts of mini-Neptunes with helium detections, TOI-836 is 8x higher in $R_X$ than TOI-2134, the least active star, but 50--100\% that of the four other mini Neptunes (TOI-560, 1430, 1683, 2076; \citealt{zhang_2022b,zhang_2023}), which have ages of a few to several hundred Myr.  

The measured X-ray luminosity can be compared to the evolutionary tracks that \cite{johnstone_2021} developed using measurements of cluster stars.  It is significantly lower than the X-ray luminosities of almost all Hyades members with similar masses (their Fig.\,8), suggesting an age greater than 650 Myr.  Comparing to the X-ray luminosity track they plot in Fig.\,11 for 0.75\,$M_\Sun$ stars, we infer an age greater than 1.7\,Gyr.  Similarly, their Fig.\,12 (left) shows that stars with the mass of TOI-836 fall below an X-ray luminosity of $10^{28}$ erg/s at around 2\,Gyr.  

Combining the gyrochronological age and the X-ray evolutionary tracks, we infer an age of $\sim$2-3 Gyr for TOI-836.

\subsection{Transit times}
TOI-836c exhibits transit timing variations, but b does not \citep{hawthorn_2023}.  For planet b, we adopt the linear ephemeris of \cite{hawthorn_2023}: $P=3.81673$\,d, $T_0=2458599.9953$ (BJD).  For planet c, modelling the TTVs is necessary to obtain an accurate transit time for the epochs of our helium observations.  Unfortunately, the planet that is causing the TTVs has not been discovered, making it difficult to model the system from Newtonian first principles.  Fortunately, the existing transit timing measurements seem well described by a sinusoid.  We therefore collect as many timing measurements as we can, fit a sinusoid with nested sampling, and use the posteriors to calculate the transit time and associated uncertainty at the epochs of the two Keck helium observations.

Our collection of 26 transit times (Table\,\ref{table:transit_times}) consists of those reported in \cite{hawthorn_2023} (5 TESS, 1 MEarth-South, 5 LCOGT, 4 CHEOPS, 1 NGTS transits), supplemented with the ultra-precise JWST/NIRSpec transit from \cite{wallack_2024} and 10 additional LCOGT transits described below.  We flip the sign on $\delta T_c$ that \cite{hawthorn_2023} reports for the LCOGT-SSO transit of 2020-04-12, because a positive $\delta T_c$ is inconsistent with their Figure 9 and with the neighboring data points.  We additionally replace their TESS transit times with those from our own pipeline \citep{dai_2021}.

\begin{table}
    \caption{Parameters from the TTV fit for planet c.  All times are in BJD (TDB) - 2,457,000.  $T_{c1}$ and $T_{c2}$ are the predicted transit times during the helium visits.}
    \centering
    \begin{tabular}{c C}
    \hline
    Parameter & \text{Value}\\
    \hline
    $P$ (d) & 8.5954667 \pm 0.000034\\
    $T_c$ & 2656.99636_{-0.00029}^{+0.00031}\\
    $A$ (d) & -0.00999_{-0.00059}^{+0.00064}\\
    $B$ (d) & 0.0021_{-0.0026}^{+0.0024}\\
    $P_{\rm super}$ (d) & 654.3 \pm 5.5\\
    \hline
    $T_{c1}$ & 3129.7430 \pm 0.0008\\
    $T_{c2}$ & 3421.9936 \pm 0.0009\\
    \hline
    \end{tabular}
    \label{table:ttv_model_params}
\end{table}

We observed ten new transits of TOI-836c from April 2022 to April 2024 in the Pan-STARRS $z_s$ band from the Las Cumbres Observatory Global Telescope (LCOGT) \citep{Brown:2013} 1\,m network nodes at Siding Spring Observatory near Coonabarabran, Australia (SSO), South Africa Astronomical Observatory near Sutherland, South Africa (SAAO), Cerro Tololo Inter-American Observatory in Chile (CTIO), and Teide Observatory on the island of Tenerife (TEID). All observations are summarized in Table\,\ref{table:transit_times}. We used the {\tt TESS Transit Finder}, which is a customized version of the {\tt Tapir} software package \citep{Jensen:2013}, to schedule our transit observations. The 1\,m telescopes are equipped with a $4096\times4096$ SINISTRO camera having an image scale of $0\farcs389$ per pixel, resulting in a $26\arcmin\times26\arcmin$ field of view.   All images were calibrated by the standard LCOGT {\tt BANZAI} pipeline \citep{McCully:2018}, and differential photometric data were extracted using {\tt AstroImageJ} \citep{Collins:2017}. We used circular photometric apertures with radii ranging from $7\farcs4$---$9\farcs8$ that excluded all of the flux from the nearest known neighbor in the Gaia DR3 catalog (Gaia DR3 4851053994762099840) that is $13\farcs7$ northeast of TOI-836.

For the ten new LCOGT transits, we fit the photometry with a simple model, consisting of a transit modelled by \texttt{batman} \citep{kreidberg_2015} and a linear slope with either airmass or time (depending on which one minimizes $\chi^2$).  We fix $R_p/R_s$ to 0.034579, $a/R_s$ to 24.56, and inclination to 88.86 degrees, taking the median of the values \cite{wallack_2024} obtained for their JWST/NIRSpec transit (their Table\,2).  We fix the limb darkening coefficients to [0.39, 0.19], which was calculated by ExoTiC-LD for the $z_s$ bandpass using the 3D ``Stagger'' grid \citep{magic_2013}.  

\begin{figure}[ht]
  \centering \subfigure {\includegraphics
    [width=0.48\textwidth]{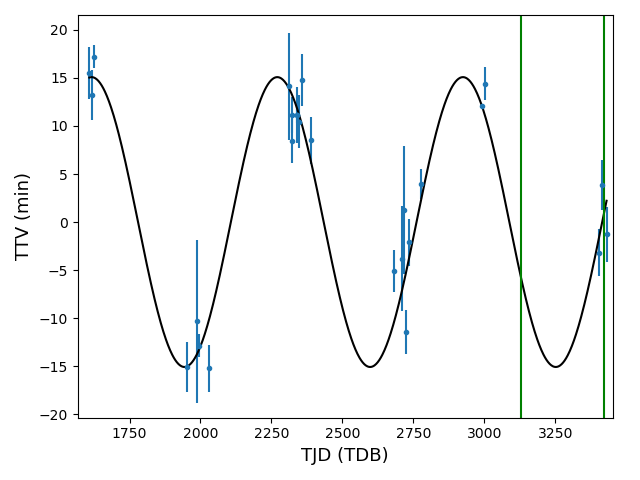}}\qquad\subfigure {\includegraphics
    [width=0.48\textwidth]{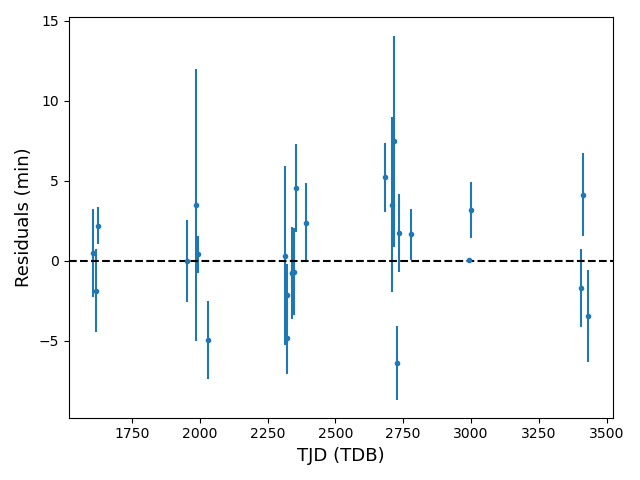}}
    \caption{Top: The transit timing variations of TOI-836c and the best-fit sinusoidal model to the TTVs.  The vertical green lines indicate the times of our two helium visits.  Bottom: the residuals of the fit.}
\label{fig:ttv_fit}
\end{figure}

After obtaining all transit times and their associated uncertainties with the help of the nested sampling package \texttt{dynesty}, we fit the following model to the transit times as a function of epoch $E$:

\begin{align*}
    T(E) &= T_0 + P\,E + \\
    & A\cos{\Big[\frac{2\pi}{P_{\rm super}} (T_0 + PE)\Big]} + B\sin{\Big[\frac{2\pi}{P_{\rm super}} (T_0 + PE)\Big]},\\
\end{align*}

where the first line is the standard linear ephemeris and the second line is a sinusoidal model for the TTVs. 
The fit and residuals are shown in Fig.\,\ref{fig:ttv_fit}.
We use \texttt{dynesty} to infer the free parameters $T_0$, $P$, $A$, $B$, and $P_{\rm super}$.  The epoch corresponding to $T_0$ was selected to minimize the covariance between $T_0$ and $P$.  We draw samples from the posterior distribution and obtain the transit times corresponding to the two helium visits, along with associated uncertainties.  The inferred parameters and transit times are all shown in Table\,\ref{table:ttv_model_params}.  TOI-836c exhibits TTVs with a semi-amplitude of $15.0\pm0.9$ minutes and a superperiod of $655\pm5$\,d.  The last zero crossing happened on TJD=$3375\pm26$ (roughly March 2024) and was in the ascending direction.  The sinusoidal TTV model constrains the transit time during the helium observations to within 70\,s.

\section{Helium observations and Data Reduction}
\label{sec:observations}
\subsection{Keck helium observations}
The observing technique and data reduction are nearly identical to that reported in \citet{zhang_2022b} and \citet{zhang_2023}.  Using the high-resolution spectrograph NIRSPEC on the Keck\,II telescope, which has a resolution of $R=32$\,k in the vicinity of the helium triplet, we took 1 minute exposures in an ABBA nodding pattern during two transits of planet c and one transit of planet b.

\begin{table}[ht]
  \centering
  \caption{Keck/NIRSPEC observations}
  \begin{tabular}{c c c c c}
  \hline
      Parameter/Visit & b & c1 & c2 \\
      \hline
      Date (UTC) & 2022-06-18 & 2023-07-04 & 2024-04-21\\
      Time (UTC) & 05:42--09:36 & 05:57--08:35 & 08:58--13:25\\
      Num. exposures & 180 & 114 & 204\\
      SNR & 170 & 120 & 160\\
      Pre-ingress (h) & 1.2 & 0 & 1.5\\
      In-transit dur. (h) & 1.8 & 1.0 & 2.5 \\
      Post-egress (h) & 1.1 & 1.6 & 0.5 \\
      Transit fraction  (\%) & 100 & 40 & 100  \\
      \hline
  \end{tabular}
  \label{table:nirspec_data}
\end{table}

The three visits are summarized in Table\,\ref{table:nirspec_data}.  The visit for planet b encompassed the entire transit, plus ample baseline on either side.  The first visit for c captured only the last 40\% of the transit, followed by 1.6\,h of post-egress baseline.  The second visit captured the whole transit in addition to 1.6\,h of pre-ingress baseline, but only 0.4\,h of post-egress baseline.

Observing conditions were decent for visits b and c2, with a seeing at the beginning of the observations of $0.6\arcsec$ and $0.65\arcsec$ respectively.  The seeing was stable during both nights, with the exception of a few minutes close to the beginning of visit b, when it was significantly degraded.  During visit c1, the seeing was $1\arcsec$ at the beginning of the night and fluctuated throughout the night.  In addition, clouds drifted in and out throughout the night.  For example, clouds blocked a substantial part of the light around 06:02 UTC and almost all of the light around 06:23 UTC, causing the loss of guiding in the latter instance.  The cloud problem was significantly ameliorated by 06:39 UTC, and for the rest of the night, clouds were never thick enough to cause Keck to lose guiding.

To analyze the Keck/NIRSPEC data, we use essentially the same methodology first introduced in \cite{zhang_2021} and used in all of our subsequent helium outflow papers, the most recent being \cite{zhang_2023}.  In brief, we construct a master dark from the pixel-wise median of 5 dark frames, each taken with a commanded exposure time of 5.5 s (true exposure time of 4.42584 s) and 20 coadds.  We construct a master flat by taking the pixelwise median of 30 flat frames, each with a commanded exposure time of 5.5 s and 20 coadds, and then subtracting the master dark from the median.  For every A$_1$B$_1$B$_2$A$_2$ nod, we compute the calibrated difference images $(A_1 - B_1)/ F$, $(B_1 - A_1) / F$, $(B_2 - A_2) / F$, and $(A_2 - B_2) / F$, where $F$ is the master flat.  We use a variant of optimal extraction adapted for curved traces, which we first introduced in \cite{zhang_2021}, to obtain the spectrum in order 70, which contains the helium triplet.  We ignore order 71 because even though it also covers the helium triplet, the triplet falls near the red edge of the order, where throughput is low and detector systematics are worse.  We also ignore the other orders on the detector.  After optimal extraction, we obtain the wavelength solution for each exposure by matching the extracted spectrum to a template generated from a PHOENIX stellar model \citep{husser_2013} and a Mauna Kea sky transmission model\footnote{\url{https://www.gemini.edu/observing/telescopes-and-sites/sites}}.

After obtaining the wavelength-calibrated spectrum for each exposure, we use \texttt{molecfit} 4.3.1 to fit the telluric absorption lines in three narrow bands (1.08590--1.086345 $\mu$m, 1.09012--1.09072 $\mu$m, 1.09245--1.09270 $\mu$m) and compute a sky transmission model across the whole wavelength range.  Dividing the spectrum by the sky transmission gives us telluric-corrected spectra.  No strong tellurics overlap the helium triplet on any of our three nights of observation (see Appendix \ref{sec:appendix_intermediate_products}).  We interpolate the telluric-corrected spectra onto a common wavelength grid, take the log of all values, and subtract off the mean of every row (time axis) and column (wavelength axis) to obtain a ``residuals grid''.  The values in the residuals grid indicate the relative change in brightness at every time and wavelength.  We subtract off changes in the continuum from the residuals grid by masking the strongly variable lines, fitting a quadratic function of wavelength for every exposure, and subtracting off the quadratic.  The plots in Appendix \ref{sec:appendix_intermediate_products} show the residuals grid from visit c2, the lines we mask, and two examples of quadratic fits.

To convert the residuals grid into an excess absorption grid, we invert the residuals grid and subtract off the mean of the out-of-transit rows for each column.  To obtain average in-transit excess absorption in the stellar frame, we take the average of the in-transit rows along the time axis.  To obtain a band-integrated light curve, we average the excess absorption within 0.75\AA{} of the main peak at 10833.32\AA{} for every row.  We also compute the average in-transit excess absorption in the planet rest frame by assuming a circular orbit with an orbital velocity of 90\,km\,s$^{-1}$, which was derived via $v_{\rm orb} = \sqrt{GM_*/a}$ using the $M_*$ and $a$ values in \cite{hawthorn_2023}.

\subsection{High-energy stellar spectrum}
\label{subsec:high_energy_spec}
Hubble Space Telescope (HST) General Observer program 16701 (PI: A. Youngblood) is a joint HST-XMM Newton program to measure the high-energy spectrum of Cycle\,1 James Webb Space Telescope (JWST) transiting planet hosts.  XMM-Newton observed TOI-836 with the European Photon Imaging Camera (EPIC) instrument for 17 ks on July 25, 2022.  Two and a half days later, HST/Space Telescope Imaging Spectrograph observed the star with the G140L, G140M, G230L, and G430L gratings. These data enable the currently unobservable Extreme Ultraviolet (EUV, $\approx100-1100$\AA{}) spectrum to be estimated with a Differential Emission Measure (DEM) model \citep{duvvuri_2021,duvvuri_2023}, which used the X-ray spectrum and FUV line fluxes as inputs.

The XMM spectrum was extracted using the standard Scientific Analysis System (SAS) routines\footnote{\url{https://www.cosmos.esa.int/web/xmm-newton/sas}}, with data from the three EPIC detectors combined in a single spectrum. The spectrum was fit with a two-temperature APEC model \citep{smith_2001, foster_2012} using the \textsc{XSPEC} package \citep{arnaud_1996}. Integrated fluxes of strong FUV lines and line groups (e.g., Si II 1300\,\AA, C II 1335\,\AA, Si IV 1400\,\AA) were measured from the G140L spectrum. The X-ray emission and FUV emission lines originate from different regions of the stellar atmosphere, so they are used to constrain the temperature and density structure of the atmosphere, expressed as a function of temperature (the DEM). The DEM is then combined with appropriate line lists to estimate the emission from the star at EUV wavelengths. The full DEM procedure is described in \cite{duvvuri_2021,duvvuri_2023}.

\begin{figure}
    \centering
    \includegraphics[width=\linewidth]{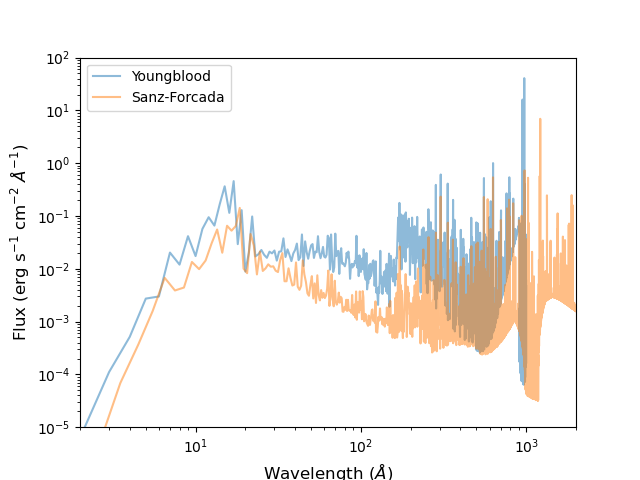}
    \caption{Two different XUV spectrum reconstructions, one by Youngblood's group following the methodology in \cite{duvvuri_2021,duvvuri_2023}, and one by Sanz-Forcada following the methodology in \citep{sanz-forcada_2025}.}
    \label{fig:XUV_comp}
\end{figure} 

Figure\,\ref{fig:XUV_comp} shows the fiducial XUV spectrum we adopt, along with an alternative reconstruction \citep{sanz-forcada_2025} using the same data but following a different methodology.  The former is higher by a factor of 3.2 in the X-ray band (5-100\,\AA{}), and higher by a factor of 4.5 in the EUV band (100--912\,\AA{}).  Possible reasons for the X-ray discrepancy--which originate mostly from the XMM-Newton data reduction--include the low counts and high background for these particular observations, and the unreliability of data below 0.3 keV.  Possible reasons for the EUV discrepancy include the differing atomic database (APEC for Sanz-Forcada, CHIANTI v10 for our reconstruction), and the different number of parameters (a smooth polynomial fit to the DEM for us; a parameter for every temperature grid point for Sanz-Forcada).  We encourage further work to explore the best way of reconstructing the XUV spectrum from XMM-Newton and HST UV data.  In the meanwhile, we adopt our own reconstruction as fiducial because the integrated 0.2--2.3 keV X-ray flux it implies, 8\e{-14} erg s$^{-1}$ cm$^{-2}$ at Earth, is similar to the value reported by the XMM-Newton Science Archive (5.8\e{-14}) as well as that reported by the eROSITA catalog (7.0\e{-14}; \citealt{merloni_2024}).

\section{Results}
\label{sec:results}

\begin{figure*}
    \centering
    \includegraphics[width=\linewidth]{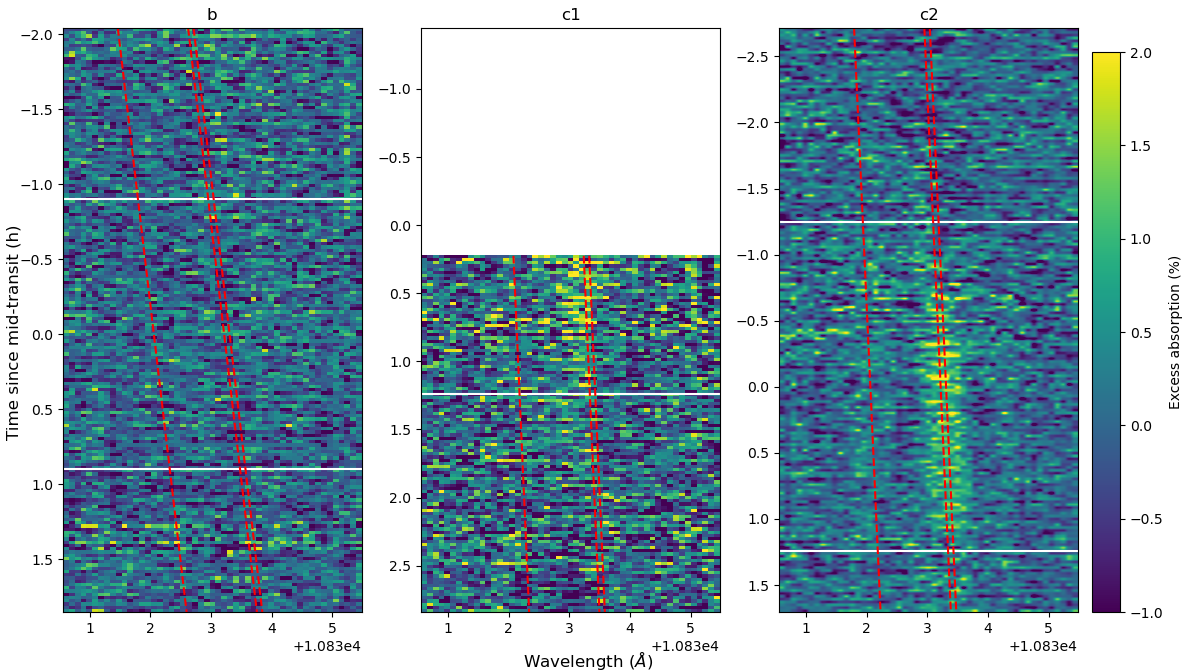}
    \caption{Excess absorption as a function of time and wavelength in the stellar frame, for planet b (left) and c (center and right).  The top and bottom white lines represent white-light ingress and egress, while the slanted red lines indicate the wavelengths of the helium triplet in the planetary frame.}
    \label{fig:excess_2D}
\end{figure*} 

Figure\,\ref{fig:excess_2D} shows the excess absorption in the three visits as a function of time and wavelength.  Planet b shows no absorption, while c shows strong absorption in both visits.  In visit c2, the absorption is asymmetric: it is higher during the half hour after egress than in the half hour before ingress.  It also appears more blueshifted in c1 than in c2.
\begin{figure}
    \centering
    \includegraphics[width=\linewidth]{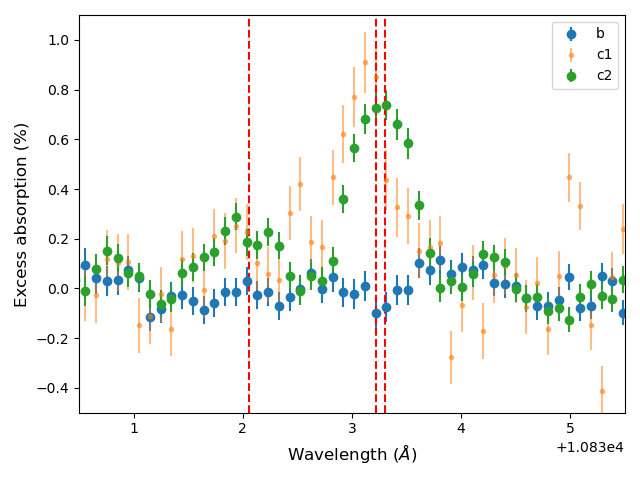}
    \caption{Average in-transit excess absorption in the planetary frame for all three visits.  The error bars are underestimated because they only include photon noise, read noise, and flatfielding error, not systematics.  Visits b and c2 are directly comparable because they both include a full transit and baseline on both sides, but c1 is not directly comparable because it is a partial transit.}
    \label{fig:absorb_1D}
\end{figure}

Figure\,\ref{fig:absorb_1D} shows the average in-transit excess absorption spectrum for both planets.  Using the nested sampling package \texttt{dynesty}, we fit c's spectrum with two Gaussians.  The Gaussians share the same standard deviation and are separated by the spacing between the two helium peaks (10833.27 - 10832.06 \AA{}= 1.21\AA{}), but the ratio between their amplitudes is allowed to vary between 1/8 and 1 (corresponding to the optically thin and thick extremes respectively).  The free parameters are the shared redshift, the peak ratio, the amplitude of the larger peak, the shared standard deviation, and an error multiple that accounts for underestimated error bars.  When calculating the likelihood, we account for covariances between data points that arise from interpolating the spectra onto a common wavelength grid and from transforming into the planetary frame (see \citealt{zhang_2021} for more details).

Our nested sampling posteriors for c are tabulated in Table\,\ref{table:gaussian_fits}.  For visit c2, there is no detectable redshift.  The peak ratio is notably significantly higher than 1/8, the value for a perfectly optically thin outflow, and much lower than 1, the value for an opaque disk.  The combined equivalent width of both lines, $\sqrt{2\pi}A\sigma(1 + r)$ for two Gaussians, is $6.7 \pm 0.6$m\AA{}.  We performed the same calculations for visit c1, but because it does not cover the entire transit, the results are not directly comparable.  To make them more comparable, we remove in-transit observations from c2 that have orbital phases not covered by c1 (namely from the beginning of the transit to 0.23\,h after mid-transit), creating a truncated dataset, c2-match.  The amplitude, peak ratio, redshift, FWHM, and equivalent width of c2-match are shown in 
Table\,\ref{table:gaussian_fits}, 
and can be compared directly with c1.  All parameters are fully consistent except the redshift, which is $0.14\pm0.88$\,km\,s$^{-1}$ for c2-match but $-4.6\pm1.6$\,km\,s$^{-1}$ for c1 -- a 2.6$\,\sigma$ discrepancy.  The bluer absorption from c1 can be seen visually in Figures \ref{fig:excess_2D} and \ref{fig:absorb_1D}.  This variability could be due to changes in the outflow or in the stellar wind.  As we show in the next subsection, it cannot be due to the radial velocity variations induced by the same planet which causes c's transit timing variations.

Using the average in-transit excess absorption spectrum, we put constraints on b by fixing the FWHM and peak ratio to the median values for c, fixing the redshift to 0, and allowing only the amplitude and error multiple to vary.  We obtained a 95\% upper limit on the amplitude of 0.047\%, corresponding to an equivalent width upper limit of 0.31 m\AA{}.  This stringent upper limit is more than 20 times lower than the measured value for c.

\begin{figure*}[ht]
  \centering \subfigure {\includegraphics
    [width=0.48\textwidth]{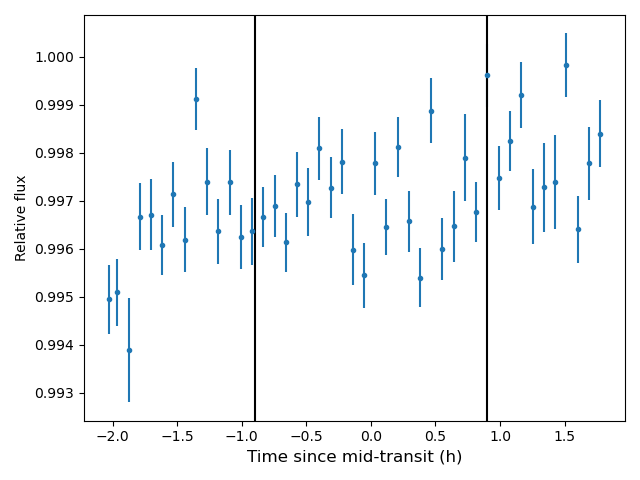}}\qquad\subfigure {\includegraphics
    [width=0.48\textwidth]{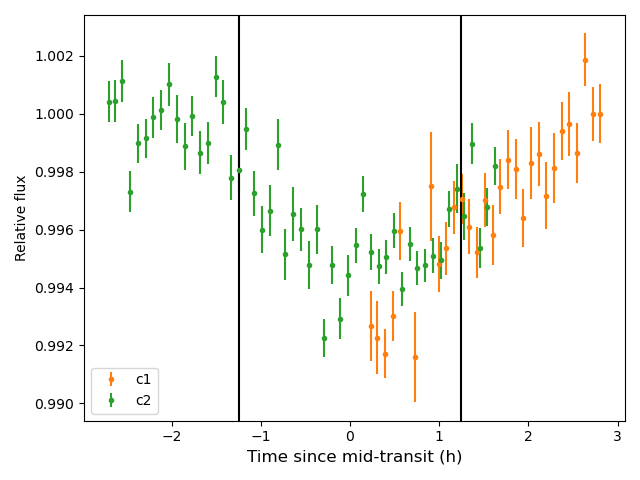}}
    \caption{Band-integrated light curves for planet b (left) and c (right).  The band we adopt is $10833.27\pm0.75$\,\AA{}, covering the main helium peak.  The data are binned by a factor of 4 for all visits.}
\label{fig:lcs}
\end{figure*}

Figure\,\ref{fig:lcs} shows the band-integrated light curve for all three visits.  Planet b shows no in-transit absorption, but there is a slight upward trend in flux, possibly due to stellar variability.  Planet c shows strong in-transit absorption in both visits.  This absorption appears to be asymmetric, with significant post-egress absorption but lower (if any) pre-ingress absorption.  To confirm this qualitative impression, we use nested sampling to fit the light curve with an opaque disk transit model.  The free parameters are the transit mid-point, transit depth, offset of the first visit relative to the second, and a separate error multiple for each night.  We obtain a bimodal posterior for the transit mid-point, with a dominant mode at $\Delta t_0=0.40\pm0.06$\,h after the white light mid-transit and a sub-dominant mode at $\Delta t_0=0.10\pm0.06$\,h.  The dominant mode corresponds to a first-visit offset close to zero, while the sub-dominant mode corresponds to an upward offset of 0.1\%. Neither mode fits the light curve shape well, likely because an opaque disk results in abrupt ingresses and egresses while the smoothness of the real outflow profile results in the gradual ingress and egress seen in the light curve.  Nevertheless, to the extent that our simple experiment can be trusted, it indicates that the light curve prefers a late egress at 2--3\,$\sigma$.

\begin{table*}[ht]
  \centering
  \begin{tabular}{c C C C C }
  \hline
    & \text{b} & \text{c2} & \text{c1} & \text{c2-match}\\
    \hline
    Amplitude (\%) & <0.047 & 0.78 \pm 0.07 & 0.74 \pm 0.14 & 0.86 \pm 0.09\\
    Ratio & 0.33^* & 0.33 \pm 0.07 & 0.27 \pm 0.14 & 0.23 \pm 0.09\\
    Redshift (km\,s$^{-1}$) & 0^* & -0.14 \pm 0.70 & -4.6 \pm 1.6 & 0.14 \pm 0.88\\
    FWHM (\AA{}) & 0.61$^*$ & 0.61 \pm 0.02 & 0.73 \pm 0.16 & 0.60 \pm 0.06\\
    EW (m\AA{}) & < 0.31 & 6.7 \pm 0.6 & 7.3 \pm 1.4 & 6.7 \pm 0.8 \\    
    \hline    
  \end{tabular}
  \caption{Double-Gaussian fit to observations.  The fiducial results for planet c are in Column\,c2, because this visit includes the complete transit.  c2-match includes only the in-transit phases also covered by c1, for comparability.  For b, parameters marked by $^*$ are fixed.}
  \label{table:gaussian_fits}
\end{table*}

\subsection{The TTV-inducing planet causes minimal RV variations}
To estimate whether the planet that induces c's TTVs can be responsible for the 4.5 km/s difference in the helium signal's blueshift between the two visits, we use the analytic expressions in \cite{lithwick_2012}.  We assume that the TTVs are induced by an outer companion that lies near (but not in) a 3:2 mean-motion resonance with c, but the calculations are correct to order-of-magnitude even for other period ratios.

The 655 d superperiod implies $|\Delta|=0.0066$ (their Eq 7).  The analytic TTV equations in \cite{lithwick_2012} are valid if the planets are not locked in resonance, a condition which is satisfied if e$_{\rm free} \lesssim \Delta^2/\mu$ (their Appendix A.2) where $\mu=4.2 \times 10^{-5}$ is the inner-planet-to-star mass ratio.  This expression evaluates to e$_{\rm free} \lesssim 1$, which is certainly satisfied.

Their Equation 8 gives the TTVs induced by the outer planet as a function of both planets' free eccentricities,  $\Delta$, and the outer-planet-to-star mass ratio $\mu'$.  Assuming the free eccentricity of both planets is not much greater than $\Delta$, we obtain that 15 min TTVs corresponds to $\mu' \sim 3.3 \times 10^{-5}$, corresponding to 7.5 $M_\Earth$ (a typical mini-Neptune mass).  If free eccentricity \textit{is} much greater than $\Delta$, we obtain a smaller mass.  This mass induces a forced eccentricity of $z_{\rm forced} \sim 0.0029$ (their Eq 13), which in turn induces radial velocity perturbations of $v_{\rm orb} z_{\rm forced} \sim 0.26$ km/s.  This is 17 times smaller than the 4.5 km/s discrepancy between the two helium line blueshifts.  We therefore conclude that radial velocity perburbations caused by the TTV-inducing planet are not responsible for the discrepancy.

A much simpler order-of-magnitude way to arrive at the same conclusion is to recall that for an eccentric orbit, the eclipse is delayed by $\delta t = \frac{2P}{\pi} e \cos{\omega}$ relative to $T_{\rm transit} + P/2$ \citep{winn_2010}, while the radial velocity of the planet during transit is $v_r = v_{\rm orb} e\cos{\omega}$.  In other words, eccentricity is roughly the orbit's fractional deviation from circularity in both time and velocity.  Equating $e \cos{\omega}$ in both expressions gives $v_r = v_{\rm orb} \frac{\pi \delta t}{2P} = 0.17$ km/s.  

\section{Modelling}
\label{sec:modelling}

\subsection{An approximate mass loss rate}
\label{subsec:mass_loss_rate}
As in our previous work (e.g. \citealt{zhang_2023b}), we estimate the mass loss rate in an order-of-magnitude way using two approaches.  The first assumes that the outflow is optically thin, that it expands at 
10\,km\,s$^{-1}$, that 25\% of the outflowing mass is in helium, and that $10^{-6}$ of the helium atoms are in the triplet ground state.  Under these assumptions, we derive in \cite{zhang_2022b} that:

\begin{align} 
     \dot{m}_{\rm obs} &= \frac{R_* \, m_e \, m_{He} \, c_s \, c^2 \, W_{\rm avg}}{0.25\,f \, e^2 \, \lambda_0^2 \, \sum_l{g_l\, f_l}}\\
     & = (3.5 \times 10^{10} \text{g/s}) \frac{R_*}{R_\Sun} \frac{W_{\rm avg}}{10\, \text{m\AA}} \frac{c_s}{\text{10\,km/s}} \frac{10^{-6}}{f}\\
     & = (0.18 M_\Earth/\text{Gyr}) \frac{R_*}{R_\Sun} \frac{W_{\rm avg}}{10\,\text{m\AA}} \frac{c_s}{\text{10 km/s}} \frac{10^{-6}}{f}\\     
     \label{eq:obs_rate}
\end{align}

For TOI-836c, we obtain a mass loss rate of 0.085 $M_\Earth$\,Gyr$^{-1}$, sufficient to strip a 1\% mass fraction envelope in 1.1\,Gyr.  For b, we obtain a 2$\sigma$ upper limit of 0.004\,$M_\Earth$\,Gyr$^{-1}$.

Our second approach to estimating the mass loss rate is a Bayesian approach that incorporates the Parker wind \citep{parker_1958} model of \cite{oklopcic_2019}, the free parameters being the mass loss rate, temperature of the isothermal outflow, and redshift.  The Parker wind code calculates the velocity profile of the outflow and the density profile of triplet ground state helium given the temperature, mass loss rate, and stellar spectrum.  These velocity and density profiles are then used to compute the excess absorption spectrum as a function of both time and wavelength, enabling comparison to the data shown in Figure\,\ref{fig:excess_2D}.  In computing the model spectrum, we assume that NIRSPEC has a spectral resolution of 32,000 and a Gaussian line spread profile.  The nested sampling run gives us a mass loss rate of $0.13\pm0.02\,M_\Earth$\,Gyr$^{-1}$ (similar to our previous order-of-magnitude estimate of 0.085), temperature of $4700\pm300$\,K, and redshift of $-0.17\pm0.27$\,km\,s$^{-1}$.  If we add the time of transit as a free parameter to account (in a physically unrealistic way) for the asymmetric light curve, we obtain a mass loss rate of $0.13_{-0.03}^{+0.05}\,M_\Earth$/Gyr, a temperature of $4700\pm600$\,K, a redshift of $0.44\pm0.58$\,km\,s$^{-1}$, and a transit midpoint of $0.21\pm0.08$\,h after the white-light mid-transit.  Uncertainties on the model parameters are almost certainly much larger than the statistical errors we quote above.  {Some sources of model error are discussed in Subsection \ref{subsec:wind_confinement}.  Another possible source of error is the H/He ratio.  If it is substantially higher than the 9:1 that both we and \cite{oklopcic_2019} adopt, as some studies of other planets suggest (e.g. \citealt{lampon_2021b}), the actual mass loss rate would be higher than estimated.

At these mass loss rates, TOI 836c would have lost 2--4\% of its mass over a lifetime of 2--3 Gyr--a very substantial fraction of the envelope of a mini-Neptune.  However, mass loss rates were unlikely to have been the same in the past due to higher XUV fluxes and a more inflated radius.  \cite{king_2020} find that for stars with TOI 836's B-V color, EUV flux declines as $t^{-0.63}$.  The lifetime-integrated EUV flux is $F_{\rm curr} t_{\rm age} / (1 - 0.63)$, so if the mass loss efficiency did not change and the X-ray contribution is neglected, planet c would have lost 5--11\% of its mass instead of 2--4\%.  However, as we shall see, higher XUV irradiation likely corresponds to lower efficiency.  If efficiency scales as $F_{XUV}^{0.5}$, the mass loss rate would scale as $t^{-0.5 * 0.63}$, resulting in an integrated mass loss only 1.5x higher than the naive estimate instead of 2.7x higher.

\subsection{pyTPCI}
\label{subsec:pyTPCI}
To interpret the observational results, we compare them to pyTPCI simulations.  pyTPCI, which will be fully described in \cite{rosener_2025}, was inspired by The PLUTO-CLOUDY Interface \citep{salz_2015}.  Like TPCI, pyTPCI is a wrapper around two long-established codes: the hydrodynamics code PLUTO 4.4 \citep{mignone_2007} and the spectral synthesis code Cloudy \citep{chatzikos_2023}.  Cloudy takes a stellar spectrum and profiles of temperature, density, and velocity, and computes population levels (including that of the helium triplet ground state) by accounting for photoionization/recombination, collisional excitation/de-excitation, atomic line heating/cooling, and advection of species, among other physical processes.  Cloudy also computes heating and cooling rates, which are fed into PLUTO for hydrodynamic evolution.  After evolving the fluid by an arbitrary 10\% in density, PLUTO shares its pressure, density, and velocity profiles with pyTPCI.  pyTPCI uses the pressure, density, and mean molecular weight from the previous Cloudy run to calculate a temperature profile, and feeds the profiles of temperature, density, and velocity back into Cloudy, completing the loop.  The main improvements of pyTPCI over TPCI are that the wrapper is entirely in Python, making it easier to modify; we change the way PLUTO handles heating, making pyTPCI more robust and less prone to crashing; and we update PLUTO and Cloudy to their latest versions, which benefit from a decade's worth of code and data improvements in addition to being much easier to compile on modern computers.  

Our simulation setup is similar to that in \cite{rosener_2025}.  It is spherically symmetric, and spans 1 to 30 planetary radii in a non-uniform grid that has larger spacings at larger radial coordinates.  To crudely account for the non-spherical nature of the outflow, we set an irradiation angle of 66 degrees, an angle which \cite{johnstone_2018} found works well for approximating Earth's globe-averaged mass loss.  We disable molecules to increase stability, set the inner boundary condition so that the temperature is the equilibrium temperature and the number density of particles is $10^{14}$ cm$^{-3}$, and evolve for 100 timesteps with advection turned off, followed by 100 timesteps with advection turned on.  Since we define one timestep (a unit time in PLUTO) to be $R_p/10\,{\rm km\,s}^{-1}$, the total simulation time is 31 hours for b and 44 hours for c.

As we showed in \cite{zhang_2022b}, the helium absorption signal depends strongly on the atmospheric metallicity.  For c, we test four different metallicities: 10x, 30x, 100x, and 200x solar.  For b, we test metallicities of 100x and 200x solar.  For each metallicity, we neglect metals with a solar number abundance smaller than $10^{-5}$, and set the abundance of the remaining metals (O, C, Ne, N, Si, Mg, Fe, S) to the metallicity times the solar abundance.  For c, 100x solar metallicity was run with both the XUV spectrum from Sanz-Forcada and from Youngblood's group; the latter has a flux 3x higher in X-rays and 4.5x higher in EUV.  The rest of the simulations were run with the Youngblood spectrum only.

\begin{figure}[ht]
  \centering \subfigure {\includegraphics
    [width=0.48\textwidth]{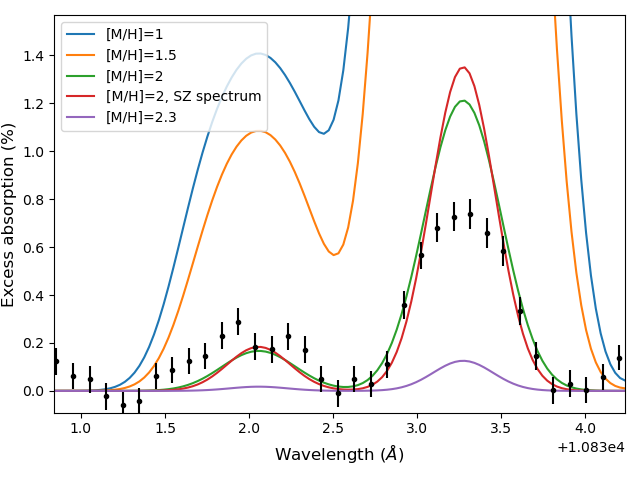}}\qquad\subfigure {\includegraphics
    [width=0.48\textwidth]{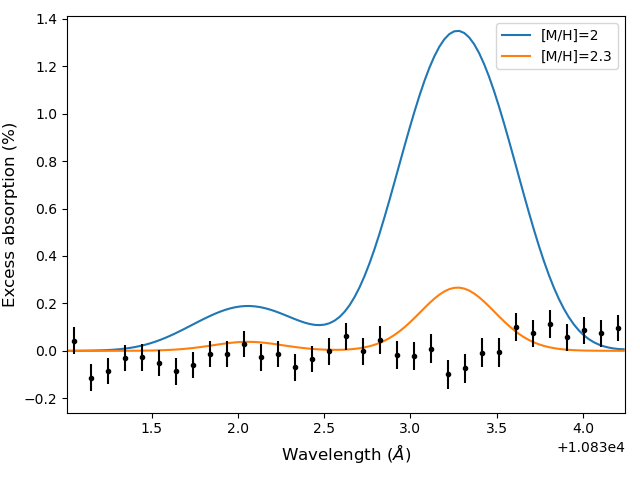}}
    \caption{pyTPCI predictions for c (top) and b (bottom) at different metallicities, compared to the observational data.  For c, we run one simulation with the XUV spectrum from David Wilson and another with the XUV spectrum from Jorge Sanz-Forcada.  Despite the factor of 3--4 flux discrepancy in both X-rays and EUV, the resulting helium absorption is remarkably similar.}
\label{fig:pytpci_models}
\end{figure}

\begin{table*}[ht]
  \centering  
  \caption{pyTPCI simulation results, after convolution with NIRSPEC's R=32k LSF. EW/A is a proxy of width, equal to 1.06(1+r)FWHM if both peaks are Gaussian.}
    \label{table:pytpci_simulation_results}
  \begin{tabular}{c C C C C C C}
  \hline
    & \text{A (\%)} & \text{Ratio} &  \text{EW/A (\AA{})} & \text{EW (m\AA{})} & \dot{M} (M_\Earth/Gyr) \\
    \hline
    b, [M/H]=2 & 1.3 & 0.14 & 0.88 & 12 & 0.35\\    
    b, [M/H]=2.3 & 0.30 & 0.17 & 0.60 & 1.8 & 0.11\\
    c, [M/H]=1   & 9.5 & 0.15 & 1.0 & 99 & 0.17\\
    c, [M/H]=1.5 & 7.6 & 0.14 & 0.91 & 69 & 0.15\\
    c, [M/H]=2 & 1.2 & 0.14 & 0.64 & 7.8 & 0.074\\
    c, [M/H]=2, SZ & 1.4 & 0.14 & 0.54 & 7.4 & 0.18\\
    c, [M/H]=2.3 & 0.12 & 0.14 & 0.50 & 0.60 & 0.017\\
    \hline    
  \end{tabular}
\end{table*}

Figure\,\ref{fig:pytpci_models} and Table \ref{table:pytpci_simulation_results} show the helium absorption predicted by these pyTPCI models, after convolution by NIRSPEC's $R=32$\,k line spread function.  For c, the peak absorption signal is an enormous 9.5\% at [M/H]=1, declining with increasing metallicity to 1.2\% at [M/H]=2 and to 0.12\% at [M/H]=2.3.  Interestingly, using the Sanz-Forcada XUV spectrum barely changes the helium signal, even though the mass loss rate it produces is 2.4x higher.  This pair of 100x metallicity models is by far the best match to the data, but they are not perfect: they over-predict the main peak but under-predict the secondary peak, indicating that the outflow is optically thicker in reality than in the models, perhaps due to confinement by the stellar wind or magnetic field.  Comparing the two planets, we see that the modelled helium signals at 100x solar metallicity look strikingly similar despite the planets' very different radii, masses, and XUV irradiation.  This is likely a coincidence, as c has a much weaker helium signal at 200x solar than b.

The dramatic collapse in the helium signal from 10x to 100x metallicity, and again from 100x to 200x metallicity, is the combination of a variety of factors that we covered in \cite{zhang_2022b}.  \cite{linssen_2024}, in their own exploration with CLOUDY of the effect of metallicity on outflows, came to similar conclusions.  The collapse is not due to the rarity of helium at high metallicities, because the number fraction of helium atoms only declines slightly within this range (7.6\% to 6.3\%).  Rather, it is due to a complex mix of factors which have the combined effect of resulting in much reduced triplet helium number densities at high metallicities -- at 10\,$R_p$, $n_{He*}$ is 18, 0.9, and 0.01\,cm$^{-3}$ at 10, 100, and 200x solar metallicities, respectively.  At high metallicity, heating becomes dominated by metal photoionization, and metal line cooling becomes dominant at less than 2 planetary radii.  The metal line cooling ``wastes'' some of the incident XUV flux, decreasing the mass loss rate.  The total emitted flux that CLOUDY predicts is 1200\,erg\,cm$^{-2}$\,s$^{-1}$ at 10x solar (31\% of the incident XUV flux), rising to 2900 (75\%) at 100x and 3820 (98\%) at 200x.  This phenomenon is reminiscent of the ``recombination-limited regime'' identified by \cite{murray-clay_2009} and described in more detail by \cite{lampon_2021a}.  Because most of the incident flux is re-radiated, mass loss is inefficient; and because a typical particle sits around repeatedly being ionized and recombining before ever escaping, the outflow becomes highly ionized at much lower radius.  While at 10x metallicity, 80\% of the helium at 10 $R_p$ is singly ionized, this fraction falls to 55\% and 16\% for 100x and 200x metallicity respectively (the rest of the helium is doubly ionized).  Since triplet ground state helium is primarily produced when singly ionized helium recombines, it is rare in the highly ionized outflow at high metallicities.
 
Another factor decreasing the mass loss rate is the lower number of easily ionizable electrons per unit mass at high metallicity: if we define ``easily ionizable'' as having an ionization energy less than or equal to the second ionization energy of helium, H has 1 easily ionizable electron per amu, He has 0.5, C has 0.25, O has 0.125, Ne has 0.1, and Fe has 0.053.  Higher metallicity atmospheres therefore have fewer easily ionizable electrons per unit mass, leading to less heating per unit mass; and since they also have higher mean molecular weights (3.85 g/mol at 200x solar, vs. 1.39 g/mol at 10x solar), even the same mass loss rate would translate to a lower particle density. 
 As a consequence $n_e$ at 10\,$R_p$ plummets from $3 \times 10^6$ at 10x solar, to $8 \times 10^5$ at 100x and $10^5$ at 200x.  The decreased availability of free electrons has another, more direct effect on the number density of triplet ground state helium: since triplet ground state helium is mostly produced by recombination, fewer free electrons means a lower number density of triplet ground state helium.  

In short, the helium signal collapses at high metallicity because both $n_e$ and $n_{\rm HeII}$ plummet.  They plummet both because there are fewer particles in the outflow in total, and because even in relative terms, electrons and singly-ionized helium become rarer.  The former is because metal lines radiate away most of the incident XUV flux, because metals are harder to fully ionize than H/He (and therefore harder to heat per unit mass via photoionization), and because there are fewer particles per unit mass.  Electrons are rarer in relative terms because metals cannot be fully ionized.  Singly-ionized helium is rarer because in the recombination limit, helium becomes doubly ionized close to the planet.

\subsection{Diffusive separation is unimportant}
\label{subsec:diffusive_sep}

pyTPCI does not compute diffusive separation, which is expected to be important when the mass loss rate is sufficiently low (c.f. \citealt{modirrousta-galian_2024,owen_2019,hu_2015}).  To estimate its importance, we compute the diffusion-limited escape rate and compare it to the actual escape rate.  If the latter is much bigger than the former, diffusive separation is unimportant.  The diffusion-limited particle flux is (c.f. \citealt{hu_2015}):
\begin{align}
    \phi_{DL} = \frac{G\,M_p \,(m_X - m_H)\,b}{R_p^2 \,k_B \,T},
\end{align}
where $m_X$ is the mass of a minor species and $b$ is the binary diffusion coefficient between hydrogen and the minor species.
\cite{zahnle_1986} catalogs the binary diffusion coefficient between hydrogen and several species in their Table\,1, including He, O, Ne, and Ar.  All of these rates are around $\sim \text{several} \times 10^{17}\, T^{0.73}$\,cm$^{-1}$\,s$^{-1}$, with the exponent varying between 0.6 and 0.75.  Plugging the $T$ scaling into the equation above, we see that $\phi_{DL}$ is a weakly decreasing function of temperature.  Using the diffusion coefficient between H and O ($4.8 \times 10^{17} \,T^{0.75}$\,cm$^{-1}$\,s$^{-1}$), we get:

\begin{align}
\frac{\phi_{DL} } { 10^{13} \text{cm}^{-2} \text{s}^{-1} } = 1.5 \, 
\Big(\frac{M_p}{M_\Earth} \Big) \Big(\frac{T}{10^3\,K}\Big)^{-0.25} \Big(\frac{R_p}{R_\Earth}\Big)^{-2} 
\ .
\end{align}

To convert this particle flux into a mass loss rate $\Phi_{DL}$, we multiply by the surface area of the planet ($4\,\pi\, R_p^2$) and by the mean molecular weight $\mu$ (roughly 1--3, depending on metallicity and ionization state), obtaining:

\begin{align}
\Phi_{DL} = (1.3 \times 10^8 \text{g\,s}^{-1}) \Big(\frac{M_p}{M_\Earth} \Big) \Big(\frac{T}{1000 K}\Big)^{-0.25} \mu
\end{align}

This equation gives roughly $10^9$\,g\,s$^{-1}$ for TOI-836c, and $6 \times 10^8$\,g\,s$^{-1}$ for TOI-836b.  By comparison, the mass loss rates we estimated in Subsection\,\ref{subsec:mass_loss_rate} from the observations of c are close to $2 \times 10^{10}$\,g\,s$^{-1}$, 20 times larger than the diffusion limited rate.  The 100x solar metallicity pyTPCI simulation, which best matches the data, has a mass loss rate of $7.5 \times 10^9$\,g\,s$^{-1}$.  Our estimate of $\Phi_{DL}$ is very conservative because we assumed a temperature of 1000\,K, whereas most of the outflow is $>2000$\,K; and we assumed a fully neutral outflow, whereas the true outflow ionizes.  The effective binary diffusion coefficient in an ionized outflow dramatically increases because ion-ion, ion-electron, and even ion-neutral interactions are substantially stronger than neutral-neutral interactions.  We conclude that diffusive separation is likely unimportant for TOI-836c.  Repeating the calculation, we find that it is also likely unimportant for the young mini-Neptunes with helium detections (TOI 560b, 1430b, 1683.01, 2076b).  For TOI 2134b, diffusive separation may be marginally significant because the star is inactive and the inferred mass loss rate is only 5 times the diffusion limit.

\subsection{The possibility of non-primordial H:He ratios}
Highly related to the question of diffusive separation is the question of the H/He ratio: is it the primoridial 9:1, as assumed by this paper and many others (e.g. \citealt{oklopcic_2018})?  By combining Ly$\alpha$ and helium triplet observations with 1D models, \cite{lampon_2020,lampon_2021b} found that HD 209458b, HD 189733b, and GJ 3470b all have elevated H/He ratios.  Modelling Ly$\alpha$ absorption is an extremely difficult problem because the observed signal is primarily controlled by the stellar tidal field, and is only logarithmically sensitive to the mass loss rate \citep{owen_2023}.  3D simulations are generally required to match observations \citep{schreyer_2024}.

An elevated H/He ratio could arise by diffusive separation: the lighter hydrogen escapes while the heavier helium stays behind.  However, we showed in the previous sub-section that diffusive separation is likely unimportant for this system.  \cite{schulik_2024} investigated the effect of fractionation on the helium signal and found that metastable state (He 2$^3$S) helium fractionates less than ground state helium due to its origin as a recombination product, giving it some of the momentum of He+ ions.  Nevertheless, they found that HD 189733b analogues orbiting M dwarfs at 0.1 AU would experience dramatic fractionation, almost completely suppressing the helium signal.  They additionally found that K2 hosts trigger significantly stronger fractionation compared to M2 or G2 hosts.  However, as they point out, the high gravity of HD 189733b means their simulations represent cases where fractionation is maximized.  They did not simulate mini-Neptunes or lower-gravity hot Jupiters.

To explore the effects of a very high H:He ratio--99:1 instead of the primordial 9:1--we run a pyTPCI simulation that is identical to the 100x solar metallicity run, except that the helium abundance is suppressed by a factor of 10.  We obtain a peak excess absorption of 0.20\% and equivalent width of 0.94 m$\AA{}$, a factor of 6--8 smaller than what we obtained with the primordial ratio.  The mass loss rate is 0.049 $M_\Earth$/Gyr, 2/3 of the mass loss rate we obtained with the primordial H/He ratio.  In addition, we re-ran the Bayesian Parker wind model of \cite{oklopcic_2019} using the 99:1 H:He ratio.  We obtain a mass loss rate of 0.48 $M_\Earth$/Gyr and temperature of 3000 K, substantially higher and colder than our result with the primordial ratio (0.13 $M_\Earth$/Gyr, 4700 K).

In the future, further observational constraints can be put on the H/He ratio with 3D models of both Ly$\alpha$ and helium absorption, or with observations of metal lines \citep{linssen_2024}.  The latter would probe the H/He ratio indirectly, by constraining the amount of fractionation in the outflow as well as the temperature and velocity profiles.

\subsection{Wind confinement and other breakdowns of radial symmetry may be important}
\label{subsec:wind_confinement}

In calculating the helium signal from the pyTPCI model, we assumed that the outflow is spherically symmetric all the way to the edge of the star.  If a strong magnetic field or stellar wind suppresses the outflow, the observations for TOI-836c could be consistent with low metallicity.  We show in Figure\,\ref{fig:c_models_vs_data_confined} the effect of truncating the outflow arbitrarily at 5 $R_p$ for planet c.  The absorption signal drops significantly at all metallicities, such that the 10x solar model most closely matches the data.  This truncation increases the effective optical depth of the outflow, increasing the secondary-to-primary peak ratio and making the predicted ratios more consistent with the data.  Truncation significantly overestimates the decrease in helium absorption due to confinement because a confined outflow does not disappear; the metastable helium is still there, still passing in front of the star, and still contributing to absorption.  With a denser outflow, reaction rates also change.  \cite{macleod_2022} found with 3D simulations that strong stellar wind confinement \textit{increases} the helium absorption by a factor of three for their particular set of parameters.

\begin{figure}
    \includegraphics[width=0.5\textwidth]{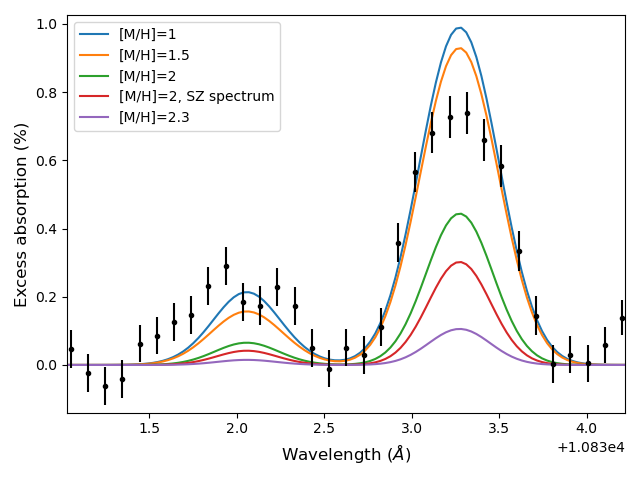}
    \caption{Predicted pyTPCI absorption for planet c, assuming arbitrarily that the outflow has a hard outer cutoff of 5 $R_p$, and that all the metastable helium disappears at higher radius.  This is likely a severe overestimation of the effects of confinement by magnetic fields or stellar winds, because a confined outflow does not disappear and should still absorb in the helium triplet.}
    \label{fig:c_models_vs_data_confined}
\end{figure}

To see whether confinement to 5\,$R_p$ is realistic, we compare the ram pressure at 5\,$R_p$, $\frac{1}{2}\rho_p\,v_p^2$, to the ram pressure of the stellar wind $\frac{1}{2}\rho_*\, v_*^2$ as well as to the magnetic pressure $B^2/(8\pi)$.  At 10x solar metallicity, $\rho_p=$4\e{-17}\,g/cm$^3$ and $v_p=9.6$\,km\,s$^{-1}$, so the ram pressure is 1.8\e{-5}\,dyne\,cm$^{-2}$.  If the stellar wind speed is 300\,km\,s$^{-1}$ (similar to the Sun), the stellar mass loss rate would need to be 2\e{13}\,g\,s$^{-1}$ to place the bow shock at 5\,$R_p$.  This mass loss rate is 12 times the solar value, or 27 times the solar value if normalized by the stellar surface area.  Whether a stellar wind this strong is realistic or not is uncertain.  $\dot{M_*}/R_*^2 = 27 \,\dot{M_\Sun}/R_\Sun^2$ would be high for the star's Rossby number of 0.8 based on the numerical simulations of \cite{chebly_2023} (see their Figure\,6), but not based on astrospheric Ly$\alpha$ observations of \cite{wood_2021} -- which, however, are subject to large uncertainties on the topology of the local ISM and on the assumed stellar wind velocity.  If a magnetic map of the star can be obtained via Zeeman-Doppler imaging, the wind can be simulated (c.f. \citealt{bellotti_2023,vidotto_2023}, where this was done for GJ 436).

Moving on to magnetic fields, a magnetic field strength of 0.02\,G is necessary for magnetic pressure to match the ram pressure at 5\,$R_p$.  The Sun's nearly-radial magnetic field at c's orbital distance, as measured by the FIELDS instrument on the Parker Solar Probe \citep{bale_2016}, was 0.0035--0.0070\,G immediately before and after the perihelion passes on 2023 September 27 and June 22, but TOI-836 could have a much stronger magnetic field.  Earth's surface magnetic field is 0.25--0.65\,G; if it were a dipole, the field would fall to 0.002--0.005\,G at $5\,R_\oplus$.  Neptune's magnetic field is weaker, with a dipole moment of 0.14 G $R_N^3$, where $R_N$ is the planet's radius \citep{connerney_1991}.  Both Earth and Neptune are very different from a tidally locked mini-Neptune like TOI-836c, and it is possible that the latter has a much stronger magnetic field.

Earlier works indicate that magnetic fields could plausibly affect the helium signal by a factor of $\sim$a few.  \cite{owen_2014,khodachenko_2015} found that a $\sim$1 G magnetic field suppresses a hot Jupiter's mass loss rate by approximately an order of magnitude, while \cite{arakcheev_2017} found that if WASP-12b had 0.1$\times$ Jupiter's magnetic moment, its mass loss rate would decrease by 70\%.  None of these papers modelled super-Earths or mini-Neptunes, nor did they model the helium signal specifically.

Aside from the stellar wind and magnetic fields, at least three other mechanisms can make the outflow non-spherical.  The first is non-uniform irradiation over the surface of the planet, which we accounted for approximately 
by adopting a 66$^\circ$ illumination angle.  The second is the Coriolis force.  A Rossby number of 1 is reached at length scales of $L=v/(2\Omega)$, where $v$ is the outflow speed.  At $v=15$\,km\,s$^{-1}$, $L=1.3\, R_\Sun$, compared to the stellar radius of 0.67\,$R_\Sun$.  The Coriolis force is therefore not dominant, but also not completely negligible.  The third mechanism is the differential stellar gravity.  The Hill sphere radius $a\,(q/3)^{1/3}$, where $q$ is the mass ratio, is $0.39\,R_\Sun$ (16\,$R_p$) for planet c; the Roche lobe 
radius $0.49\,a\,q^{2/3} / \left( 0.6\,q^{2/3} + \ln(1+q^{1/3})\right)$ is $0.27\,R_\Sun$ (12\,$R_p$).  Since none of these three mechanisms confine the outflow in the way that magnetic fields and stellar winds do, it is not clear whether or by how much they would decrease the equivalent width of the helium absorption.  We encourage 3D MHD simulations of TOI 836c to explore the impact of all these factors.

\section{Discussion and Conclusion}
\label{sec:discussion_conclusion}
\begin{figure*}
    \includegraphics[width=\textwidth]{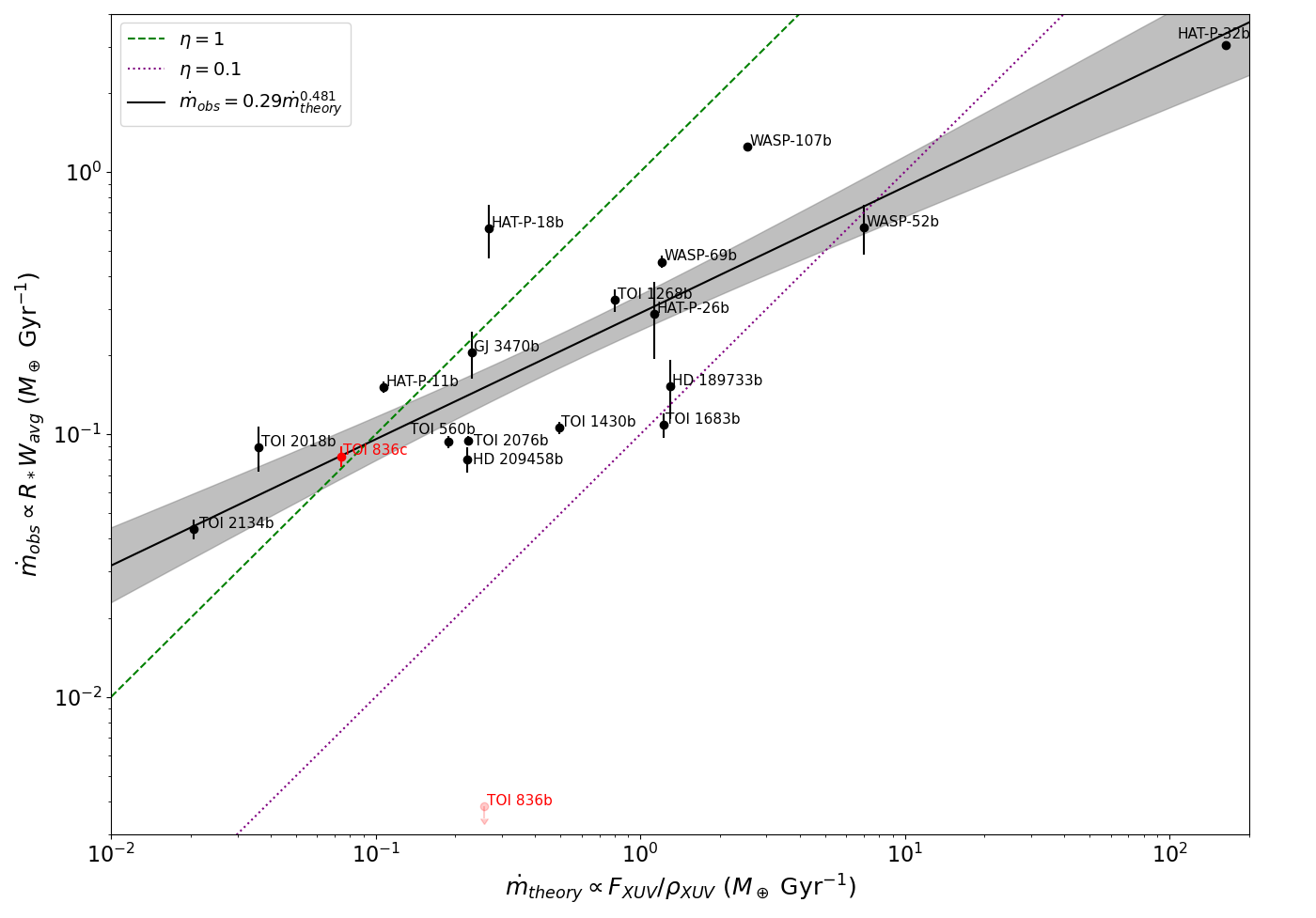}
    \caption{The correlation between the ``order-of-magnitude mass loss rate'' estimated from the equivalent width of helium absorption, and the energy-limited mass loss rate.  We first reported the correlation in \cite{zhang_2023b}.  This plot includes the same data as Figure 3 of \cite{zhang_2023b}, except that we added the detections of TOI-1268b and TOI-2018b \citep{orell-miquel_2024}, removed non-detections due to their proliferation, and added the results from this work.  Note the extreme discrepancy between TOI-836b and c.}
    \label{fig:mdot_relation}
\end{figure*}

It is clear from our observations that TOI-836b and c have very different atmospheres.  To put the non-detection from b in perspective, we plot both planets on Figure\,\ref{fig:mdot_relation} against all exoplanet helium detections.  This plot shows the correlation, first reported in \cite{zhang_2023b}, between the ``order-of-magnitude mass loss rate'' estimated from the equivalent width of helium absorption and the energy-limited mass loss rate.  Figure\,\ref{fig:mdot_relation} shows that TOI-836c perfectly follows the trendline, while b is below the trendline by almost two orders of magnitude.  This discrepancy shows that c is a mini-Neptune with at least some remaining primordial atmosphere, while b has lost its initial H$_2$/He envelope.

Helium observations have the potential not just to reveal the fundamental nature of exoplanet atmospheres, but to measure their composition.  It has been a long-standing goal of the exoplanet community to measure the composition of exoplanet atmospheres, both for its own sake and to understand planet formation and evolution.  The standard method is to obtain transmission or emission spectra of the lower atmosphere and perform a Bayesian retrieval.  However, retrievals often constrain composition very poorly, and even on JWST data of hot Jupiters over broad wavelength ranges, order-of-magnitude uncertainties on volume mixing ratios are not uncommon (see e.g. retrievals on the WASP-43b MIRI emission spectra in \citealt{bell_2024}).  The fundamental reasons that obtaining metallicity from transmission and emission spectra is challenging are the weak dependence of spectral features on metallicity (especially in the presence of clouds), and the degeneracy of metallicity with equally unknown parameters, such as haze properties and temperature/pressure profiles.

As we saw in Subsection\,\ref{subsec:pyTPCI}, the helium signal does not have the first problem: it is \textit{extremely} sensitive to metallicity between 10x and 200x solar, with the equivalent width predicted for c dropping by a factor of 13 between 10x and 100x solar, and by a further factor of 12 between 100x and 200x solar.  Helium observations therefore hold the promise of measuring atmospheric metallicity to better than 0.3 dex, if the metallicity is in the neighorhood of 100x solar.  100x solar is a scientifically significant threshold because it is the transition between a H/He dominated atmosphere and a metal-dominated atmosphere by mass.  The first JWST phase curve of a mini-Neptune suggests an atmospheric metallicity equal or greater than 100x solar \citep{kempton_2023}, and the first non-flat JWST transmission spectra of temperate mini-Neptunes are also consistent with metallicities in this neighborhood: \cite{benneke_2024} found $225_{-86}^{+98}$ for TOI-270d, while the molecular abundances reported by \cite{madhusudhan_2023} for K2-18b are consistent with 100x solar \citep{wogan_2024}.  \cite{benneke_2024} proposed a ``miscible-envelope sub-Neptune'' paradigm which they suggest includes the vast majority of known sub-Neptunes.  In this paradigm, accreted H$_2$/He reacts with the magma at the base of the envelope, chemically producing water (among other molecules).  At temperatures and pressures beyond the critical point (218 atmospheres, 647\,K), the discontinuous phase transition between liquid water and water vapor melts away into a continuum: a supercritical state.  The chemically produced water therefore does not form an ocean, but mixes into the envelope.  \cite{benneke_2024} found that even without ice accretion, these magma-envelope interactions can explain the $\sim$200x solar metallicity of TOI-270d.  If the vast majority of known sub-Neptunes are indeed ``miscible-envelope sub-Neptunes'', the vast majority could have similar metallicities.

If helium observations can reveal which mini-Neptunes have a metallicity $\gtrsim 100$x solar and which have metallicities typical of hot Jupiters ($\sim10$x), they can test the miscible-envelope paradigm and reveal the fundamental nature of one of the universe's most common class of exoplanets.  They can do so even for planets obscured by haze, and even for planets with relatively low observational favorability, since a 10\% helium absorption signal would be detectable with Keck even around a J=13 star.  While most observational efforts thus far have focused on K dwarfs because \cite{oklopcic_2019} predicted that their spectrum is optimal for populating the metastable state, this is not an absolute requirement; \cite{biassoni_2024} has predicted large (30\%) absorption signals for planets orbiting M dwarfs, and the largest ever helium signal was detected around a F dwarf (HAT-P-32; \citealt{czesla_2022}).  The question of which M dwarf planets can hold on to their atmospheres is of prime importance to exoplanet astronomy and an active current area of research.  If helium observations can elucidate the physics of mass loss from mini Neptunes orbiting M dwarfs, our understanding of the atmospheres (or lack thereof) of their super-Earth cousins could also be improved.

However, there are significant theoretical uncertainties that must be overcome before these ambitious goals can be considered met robustly, and the most important among them is understanding outflow confinement due to the stellar wind or magnetic fields (Subsection\,\ref{subsec:wind_confinement}).  We encourage 3D radiation MHD simulations of mini-Neptune outflows to determine the magnitude of these effects, and therefore the robustness of the helium absorption equivalent width to the unknown stellar wind, interplanetary magnetic field, and planetary magnetic field.  With pyTPCI and other 1D radiation hydrodynamics codes, we encourage exploration of other dimensions of composition variability, especially the C/O ratio and the helium fraction (which recent work by \cite{lampon_2020,lampon_2021b} claims are depleted in some observed outflows).  A depleted helium fraction would significantly decrease helium absorption, throwing off metallicity inferences.  On the observational side, we encourage novel tests of mass loss models.  For example, for mini-Neptunes of high metallicity, metals in the outflow can be probed directly via their absorption lines, some of which can have considerable depth even for a solar metallicity atmosphere \citep{linssen_2024}.  These measurements would allow for more stringent conclusions on outflow composition.

\appendix

\begin{table}[htbp]
    \caption{Observed transit times for TOI-836c}    
    \label{table:transit_times}
    \setlength{\tabcolsep}{6pt}
    \centering
    \begin{tabular}{c c c c c}
        \hline
        Facility & Date & TBJD$_{\rm TDB}$ & Error \\
	\hline
        TESS-S11 & 2019-05-04 & 1608.36005 & 0.00190\\
        TESS-S11 & 2019-05-13 & 1616.95390 & 0.00180\\
        MEarth-South & 2019-05-22 & 1625.55216 & 0.00081\\
        LCOGT-SSO & 2020-04-12 & 1952.15748 & 0.00178\\
        LCOGT-SAAO & 2020-05-17 & 1986.54265 & 0.00590\\
        CHEOPS & 2020-05-25 & 1995.13636 & 0.00081\\
        CHEOPS & 2020-06-29 & 2029.51660 & 0.00169\\
        LCOGT-SSO & 2021-04-08 & 2313.18741 & 0.00388\\
        NGTS & 2021-04-17 & 2321.78080 & 0.00132\\
        LCOGT-CTIO & 2021-04-17 & 2321.78080 & 0.00132\\
        CHEOPS & 2021-05-04 & 2338.97174 & 0.00062\\
        LCOGT-CTIO & 2021-06-25 & 2390.54269 & 0.00171\\
        TESS-S38 & 2021-05-04 & 2338.97174 & 0.00200\\
        TESS-S38 & 2021-05-13 & 2347.56672 & 0.00190\\
        TESS-S38 & 2021-05-21 & 2356.16520 & 0.00190\\
        LCOGT-CTIO & 2022-04-13 & 2682.77913 & 0.0015\\
        LCOGT-TEID  & 2022-05-08 & 2708.56644 & 0.0038\\
        LCOGT-SSO & 2022-05-17 & 2717.16543 & 0.0046\\
        LCOGT-CTIO & 2022-05-26 & 2725.75209 & 0.0016\\
        LCOGT-SAAO & 2022-06-03 & 2734.35401 & 0.0017\\
        LCOGT-SAAO & 2022-07-16 & 2777.33559 & 0.0011\\
        JWST & 2023-02-16 & 2992.22771 & 0.00005\\        
        LCOGT-CTIO & 2023-02-25 & 3000.82497 & 0.0012\\
        LCOGT-CTIO & 2024-04-04 & 3404.79970 & 0.0017\\
        LCOGT-SAAO & 2024-04-12 & 3413.40007 & 0.0018\\
        LCOGT-SAAO & 2024-04-29 & 3430.58745 & 0.0020\\
        \hline
    \end{tabular}
\end{table}

\section{Intermediate products from data reduction}
\label{sec:appendix_intermediate_products}

\begin{figure}[ht]
  \centering 
  \includegraphics[width=\textwidth]{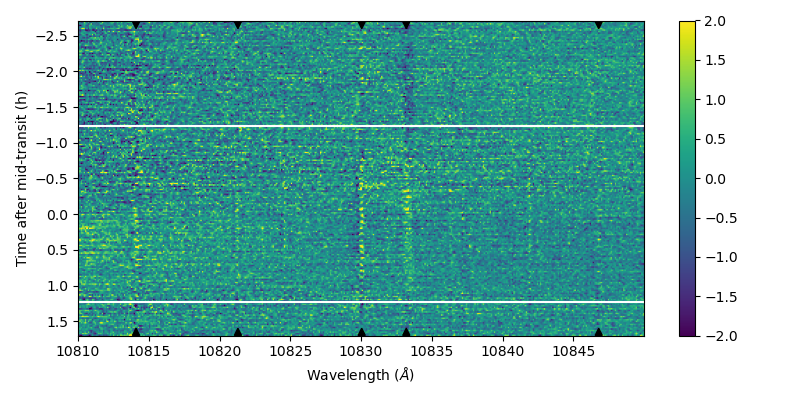}
    \caption{``Residuals grid'' for visit c2.  Colors represent deviations, in percent, from the time-averaged mean flux at that wavelength (lower flux is yellower).  Black triangles indicate the lines that we masked out before fitting a quadratic function to each row, which in turn was done to remove continuum variations.}
\label{fig:residuals_grid}
\end{figure}

\begin{figure}[ht]
  \centering 
  \subfigure {\includegraphics[width=0.49\textwidth]{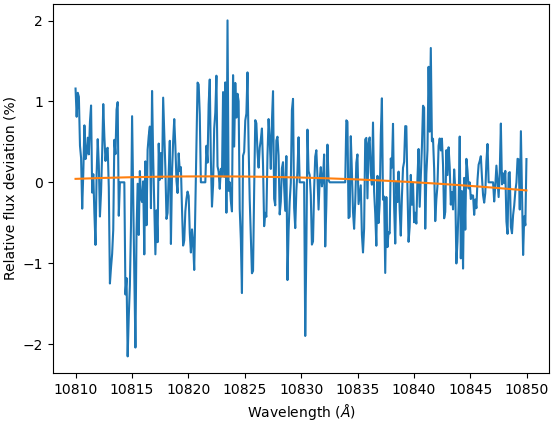}}
  \subfigure {\includegraphics[width=0.49\textwidth]{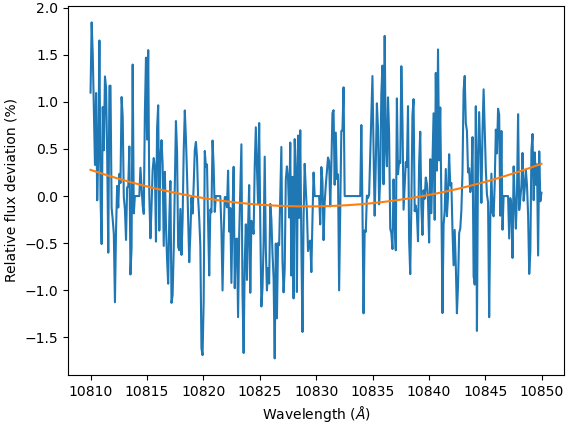}}
  \caption{Quadratic fits to the 0th and 80th rows of the ``residuals grid'' for visit c2.  They are respectively examples of spectra with particularly low and particularly severe continuum variations.}
\label{fig:quadratic_fits}
\end{figure}

\begin{figure}[ht]
  \centering 
  \subfigure {\includegraphics[width=0.48\textwidth]{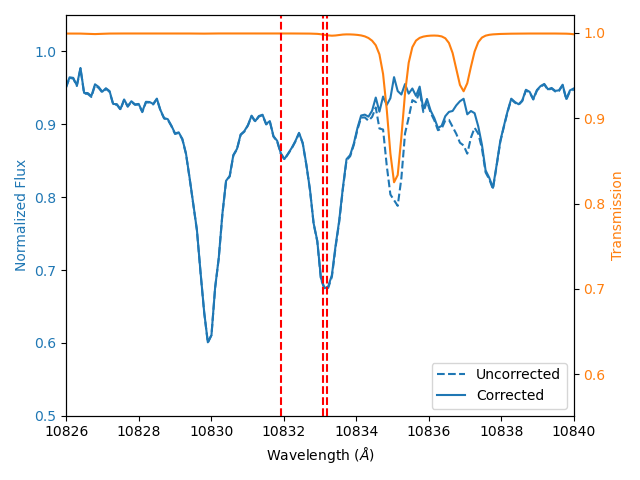}}
  \subfigure {\includegraphics[width=0.48\textwidth]{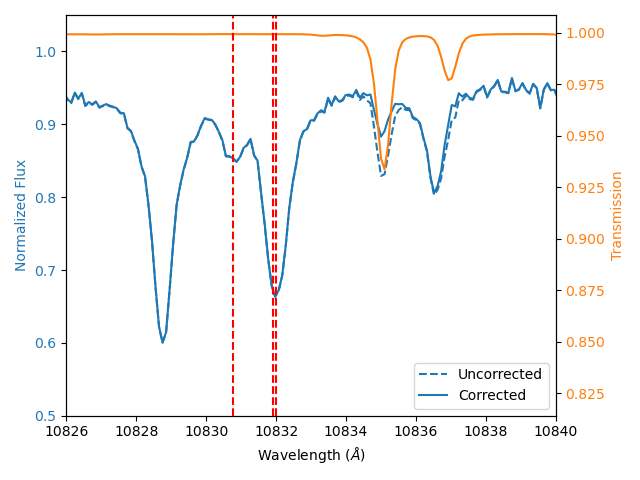}}
  \subfigure {\includegraphics[width=0.48\textwidth]{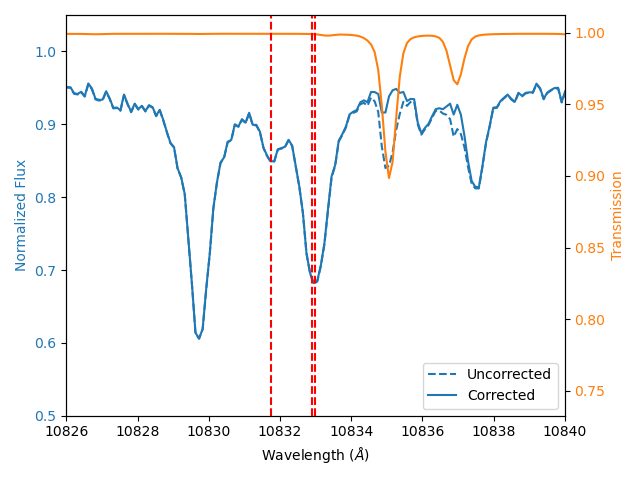}}    
    \caption{The \texttt{molecfit} telluric correction step.  For visits c1 (upper left), c2 (upper right), and b (lower center), we plot an example uncorrected spectrum, the corrected version of that spectrum, and transmission function used to make the correction. 
 The wavelengths of the helium triplet are indicated in red.}
\label{fig:spectra_and_trans}
\end{figure}

\textit{Software:}  \texttt{numpy \citep{van_der_walt_2011}, scipy \citep{virtanen_2020}, matplotlib \citep{hunter_2007}, dynesty \citep{speagle_2019}, SAS \citep{gabriel_2004}, XSPEC \citep{arnaud_1996}, molecfit \citep{smette_2015}}

\section{Acknowledgments}
We thank Jorge Sanz-Forcada for generating a XUV spectrum of TOI-836 based on the methodology of \citep{sanz-forcada_2025}.  MZ thanks the Heising-Simons Foundation for funding his 51 Pegasi b fellowship.  The postdoctoral fellowship of KB is funded by F.R.S.-FNRS grant T.0109.20 and by the Francqui Foundation.

The helium data presented herein were obtained at the W. M. Keck Observatory, which is operated as a scientific partnership among the California Institute of Technology, the University of California and the National Aeronautics and Space Administration. The Observatory was made possible by the generous financial support of the W. M. Keck Foundation.  We used observations obtained with XMM-Newton, an ESA science mission with instruments and contributions directly funded by ESA Member States and NASA.  We acknowledge funding from HST GO 17221.


This work makes use of the CHIANTI v10 database \citep{dere_1997,del_zanna_2021}.


This work makes use of observations from the LCOGT network. Part of the LCOGT telescope time was granted by NOIRLab through the Mid-Scale Innovations Program (MSIP). MSIP is funded by NSF.


This research has made use of the Exoplanet Follow-up Observation Program (ExoFOP; DOI: 10.26134/ExoFOP5) website, which is operated by the California Institute of Technology, under contract with the National Aeronautics and Space Administration under the Exoplanet Exploration Program.


Funding for the TESS mission is provided by NASA's Science Mission Directorate. KAC and CNW acknowledge support from the TESS mission via subaward s3449 from MIT.

\bibliographystyle{aasjournal} \bibliography{main}

\begin{thebibliography}{}
\expandafter\ifx\csname natexlab\endcsname\relax\def\natexlab#1{#1}\fi
\providecommand{\url}[1]{\href{#1}{#1}}
\providecommand{\dodoi}[1]{doi:~\href{http://doi.org/#1}{\nolinkurl{#1}}}
\providecommand{\doeprint}[1]{\href{http://ascl.net/#1}{\nolinkurl{http://ascl.net/#1}}}
\providecommand{\doarXiv}[1]{\href{https://arxiv.org/abs/#1}{\nolinkurl{https://arxiv.org/abs/#1}}}

\bibitem[{Alderson {et~al.}(2024)Alderson, Batalha, Wakeford, Wallack, Aguichine, Teske, Redai, Alam, Batalha, Gao, Kirk, López-Morales, Moran, Scarsdale, Wogan, \& Wolfgang}]{alderson_2024}
Alderson, L., Batalha, N.~E., Wakeford, H.~R., {et~al.} 2024, AJ, 167, 216, \dodoi{10.3847/1538-3881/ad32c9}

\bibitem[{{Arakcheev} {et~al.}(2017){Arakcheev}, {Zhilkin}, {Kaigorodov}, {Bisikalo}, \& {Kosovichev}}]{arakcheev_2017}
{Arakcheev}, A.~S., {Zhilkin}, A.~G., {Kaigorodov}, P.~V., {Bisikalo}, D.~V., \& {Kosovichev}, A.~G. 2017, Astronomy Reports, 61, 932, \dodoi{10.1134/S1063772917110014}

\bibitem[{{Arnaud}(1996)}]{arnaud_1996}
{Arnaud}, K.~A. 1996, in Astronomical Society of the Pacific Conference Series, Vol. 101, Astronomical Data Analysis Software and Systems V, ed. G.~H. {Jacoby} \& J.~{Barnes}, 17

\bibitem[{{Bale} {et~al.}(2016){Bale}, {Goetz}, {Harvey}, {Turin}, {Bonnell}, {Dudok de Wit}, {Ergun}, {MacDowall}, {Pulupa}, {Andre}, {Bolton}, {Bougeret}, {Bowen}, {Burgess}, {Cattell}, {Chandran}, {Chaston}, {Chen}, {Choi}, {Connerney}, {Cranmer}, {Diaz-Aguado}, {Donakowski}, {Drake}, {Farrell}, {Fergeau}, {Fermin}, {Fischer}, {Fox}, {Glaser}, {Goldstein}, {Gordon}, {Hanson}, {Harris}, {Hayes}, {Hinze}, {Hollweg}, {Horbury}, {Howard}, {Hoxie}, {Jannet}, {Karlsson}, {Kasper}, {Kellogg}, {Kien}, {Klimchuk}, {Krasnoselskikh}, {Krucker}, {Lynch}, {Maksimovic}, {Malaspina}, {Marker}, {Martin}, {Martinez-Oliveros}, {McCauley}, {McComas}, {McDonald}, {Meyer-Vernet}, {Moncuquet}, {Monson}, {Mozer}, {Murphy}, {Odom}, {Oliverson}, {Olson}, {Parker}, {Pankow}, {Phan}, {Quataert}, {Quinn}, {Ruplin}, {Salem}, {Seitz}, {Sheppard}, {Siy}, {Stevens}, {Summers}, {Szabo}, {Timofeeva}, {Vaivads}, {Velli}, {Yehle}, {Werthimer}, \& {Wygant}}]{bale_2016}
{Bale}, S.~D., {Goetz}, K., {Harvey}, P.~R., {et~al.} 2016, \ssr, 204, 49, \dodoi{10.1007/s11214-016-0244-5}

\bibitem[{{Bean} {et~al.}(2021){Bean}, {Raymond}, \& {Owen}}]{bean_2021}
{Bean}, J.~L., {Raymond}, S.~N., \& {Owen}, J.~E. 2021, Journal of Geophysical Research (Planets), 126, e06639, \dodoi{10.1029/2020JE006639}

\bibitem[{{Bell} {et~al.}(2024){Bell}, {Crouzet}, {Cubillos}, {Kreidberg}, {Piette}, {Roman}, {Barstow}, {Blecic}, {Carone}, {Coulombe}, {Ducrot}, {Hammond}, {Mendon{\c{c}}a}, {Moses}, {Parmentier}, {Stevenson}, {Teinturier}, {Zhang}, {Batalha}, {Bean}, {Benneke}, {Charnay}, {Chubb}, {Demory}, {Gao}, {Lee}, {L{\'o}pez-Morales}, {Morello}, {Rauscher}, {Sing}, {Tan}, {Venot}, {Wakeford}, {Aggarwal}, {Ahrer}, {Alam}, {Baeyens}, {Barrado}, {Caceres}, {Carter}, {Casewell}, {Challener}, {Crossfield}, {Decin}, {D{\'e}sert}, {Dobbs-Dixon}, {Dyrek}, {Espinoza}, {Feinstein}, {Gibson}, {Harrington}, {Helling}, {Hu}, {Iro}, {Kempton}, {Kendrew}, {Komacek}, {Krick}, {Lagage}, {Leconte}, {Lendl}, {Lewis}, {Lothringer}, {Malsky}, {Mancini}, {Mansfield}, {Mayne}, {Evans-Soma}, {Molaverdikhani}, {Nikolov}, {Nixon}, {Palle}, {Petit dit de la Roche}, {Piaulet}, {Powell}, {Rackham}, {Schneider}, {Steinrueck}, {Taylor}, {Welbanks}, {Yurchenko}, {Zhang}, \& {Zieba}}]{bell_2024}
{Bell}, T.~J., {Crouzet}, N., {Cubillos}, P.~E., {et~al.} 2024, Nature Astronomy, \dodoi{10.1038/s41550-024-02230-x}

\bibitem[{{Bellotti} {et~al.}(2023){Bellotti}, {Fares}, {Vidotto}, {Morin}, {Petit}, {Hussain}, {Bourrier}, {Donati}, {Moutou}, \& {H{\'e}brard}}]{bellotti_2023}
{Bellotti}, S., {Fares}, R., {Vidotto}, A.~A., {et~al.} 2023, \aap, 676, A139, \dodoi{10.1051/0004-6361/202346675}

\bibitem[{{Benneke} {et~al.}(2024){Benneke}, {Roy}, {Coulombe}, {Radica}, {Piaulet}, {Ahrer}, {Pierrehumbert}, {Krissansen-Totton}, {Schlichting}, {Hu}, {Yang}, {Christie}, {Thorngren}, {Young}, {Pelletier}, {Knutson}, {Miguel}, {Evans-Soma}, {Dorn}, {Gagnebin}, {Fortney}, {Komacek}, {MacDonald}, {Raul}, {Cloutier}, {Acuna}, {Lafreni{\`e}re}, {Cadieux}, {Doyon}, {Welbanks}, \& {Allart}}]{benneke_2024}
{Benneke}, B., {Roy}, P.-A., {Coulombe}, L.-P., {et~al.} 2024, arXiv e-prints, arXiv:2403.03325, \dodoi{10.48550/arXiv.2403.03325}

\bibitem[{{Biassoni} {et~al.}(2024){Biassoni}, {Caldiroli}, {Gallo}, {Haardt}, {Spinelli}, \& {Borsa}}]{biassoni_2024}
{Biassoni}, F., {Caldiroli}, A., {Gallo}, E., {et~al.} 2024, \aap, 682, A115, \dodoi{10.1051/0004-6361/202347517}

\bibitem[{{Bouma} {et~al.}(2023){Bouma}, {Palumbo}, \& {Hillenbrand}}]{bouma_2023}
{Bouma}, L.~G., {Palumbo}, E.~K., \& {Hillenbrand}, L.~A. 2023, \apjl, 947, L3, \dodoi{10.3847/2041-8213/acc589}

\bibitem[{{Brown} {et~al.}(2013){Brown}, {Baliber}, {Bianco}, {Bowman}, {Burleson}, {Conway}, {Crellin}, {Depagne}, {De Vera}, {Dilday}, {Dragomir}, {Dubberley}, {Eastman}, {Elphick}, {Falarski}, {Foale}, {Ford}, {Fulton}, {Garza}, {Gomez}, {Graham}, {Greene}, {Haldeman}, {Hawkins}, {Haworth}, {Haynes}, {Hidas}, {Hjelstrom}, {Howell}, {Hygelund}, {Lister}, {Lobdill}, {Martinez}, {Mullins}, {Norbury}, {Parrent}, {Paulson}, {Petry}, {Pickles}, {Posner}, {Rosing}, {Ross}, {Sand}, {Saunders}, {Shobbrook}, {Shporer}, {Street}, {Thomas}, {Tsapras}, {Tufts}, {Valenti}, {Vander Horst}, {Walker}, {White}, \& {Willis}}]{Brown:2013}
{Brown}, T.~M., {Baliber}, N., {Bianco}, F.~B., {et~al.} 2013, \pasp, 125, 1031, \dodoi{10.1086/673168}

\bibitem[{{Chatzikos} {et~al.}(2023){Chatzikos}, {Bianchi}, {Camilloni}, {Chakraborty}, {Gunasekera}, {Guzm{\'a}n}, {Milby}, {Sarkar}, {Shaw}, {van Hoof}, \& {Ferland}}]{chatzikos_2023}
{Chatzikos}, M., {Bianchi}, S., {Camilloni}, F., {et~al.} 2023, \rmxaa, 59, 327, \dodoi{10.22201/ia.01851101p.2023.59.02.12}

\bibitem[{Chebly {et~al.}(2023)Chebly, Alvarado-Gómez, Poppenhäger, \& Garraffo}]{chebly_2023}
Chebly, J.~J., Alvarado-Gómez, J.~D., Poppenhäger, K., \& Garraffo, C. 2023, MNRAS, 524, 5060, \dodoi{10.1093/mnras/stad2100}

\bibitem[{{Collins} {et~al.}(2017){Collins}, {Kielkopf}, {Stassun}, \& {Hessman}}]{Collins:2017}
{Collins}, K.~A., {Kielkopf}, J.~F., {Stassun}, K.~G., \& {Hessman}, F.~V. 2017, \aj, 153, 77, \dodoi{10.3847/1538-3881/153/2/77}

\bibitem[{{Connerney} {et~al.}(1991){Connerney}, {Acuna}, \& {Ness}}]{connerney_1991}
{Connerney}, J.~E.~P., {Acuna}, M.~H., \& {Ness}, N.~F. 1991, \jgr, 96, 19023, \dodoi{10.1029/91JA01165}

\bibitem[{{Czesla} {et~al.}(2022){Czesla}, {Lamp{\'o}n}, {Sanz-Forcada}, {Garc{\'\i}a Mu{\~n}oz}, {L{\'o}pez-Puertas}, {Nortmann}, {Yan}, {Nagel}, {Yan}, {Schmitt}, {Aceituno}, {Amado}, {Caballero}, {Casasayas-Barris}, {Henning}, {Khalafinejad}, {Molaverdikhani}, {Montes}, {Pall{\'e}}, {Reiners}, {Schneider}, {Ribas}, {Quirrenbach}, {Zapatero Osorio}, \& {Zechmeister}}]{czesla_2022}
{Czesla}, S., {Lamp{\'o}n}, M., {Sanz-Forcada}, J., {et~al.} 2022, \aap, 657, A6, \dodoi{10.1051/0004-6361/202039919}

\bibitem[{{Dai} {et~al.}(2021){Dai}, {Howard}, {Batalha}, {Beard}, {Behmard}, {Blunt}, {Brinkman}, {Chontos}, {Crossfield}, {Dalba}, {Dressing}, {Fulton}, {Giacalone}, {Hill}, {Huber}, {Isaacson}, {Kane}, {Lubin}, {Mayo}, {Mo{\v{c}}nik}, {Akana Murphy}, {Petigura}, {Rice}, {Robertson}, {Rosenthal}, {Roy}, {Rubenzahl}, {Weiss}, {Zandt}, {Beichman}, {Ciardi}, {Collins}, {Gonzales}, {Howell}, {Matson}, {Matthews}, {Schlieder}, {Schwarz}, {Ricker}, {Vanderspek}, {Latham}, {Seager}, {Winn}, {Jenkins}, {Caldwell}, {Colon}, {Dragomir}, {Lund}, {McLean}, {Rudat}, \& {Shporer}}]{dai_2021}
{Dai}, F., {Howard}, A.~W., {Batalha}, N.~M., {et~al.} 2021, \aj, 162, 62, \dodoi{10.3847/1538-3881/ac02bd}

\bibitem[{{Del Zanna} {et~al.}(2021){Del Zanna}, {Dere}, {Young}, \& {Landi}}]{del_zanna_2021}
{Del Zanna}, G., {Dere}, K.~P., {Young}, P.~R., \& {Landi}, E. 2021, \apj, 909, 38, \dodoi{10.3847/1538-4357/abd8ce}

\bibitem[{{Dere} {et~al.}(1997){Dere}, {Landi}, {Mason}, {Monsignori Fossi}, \& {Young}}]{dere_1997}
{Dere}, K.~P., {Landi}, E., {Mason}, H.~E., {Monsignori Fossi}, B.~C., \& {Young}, P.~R. 1997, \aaps, 125, 149, \dodoi{10.1051/aas:1997368}

\bibitem[{{Duvvuri} {et~al.}(2023){Duvvuri}, {Cauley}, {Aguirre}, {Kilgard}, {France}, {Berta-Thompson}, \& {Pineda}}]{duvvuri_2023}
{Duvvuri}, G.~M., {Cauley}, P.~W., {Aguirre}, F.~C., {et~al.} 2023, \aj, 166, 196, \dodoi{10.3847/1538-3881/acfa74}

\bibitem[{{Duvvuri} {et~al.}(2021){Duvvuri}, {Sebastian Pineda}, {Berta-Thompson}, {Brown}, {France}, {Kowalski}, {Redfield}, {Tilipman}, {Vieytes}, {Wilson}, {Youngblood}, {Froning}, {Linsky}, {Parke Loyd}, {Mauas}, {Miguel}, {Newton}, {Rugheimer}, \& {Christian Schneider}}]{duvvuri_2021}
{Duvvuri}, G.~M., {Sebastian Pineda}, J., {Berta-Thompson}, Z.~K., {et~al.} 2021, \apj, 913, 40, \dodoi{10.3847/1538-4357/abeaaf}

\bibitem[{{Foster} {et~al.}(2012){Foster}, {Ji}, {Smith}, \& {Brickhouse}}]{foster_2012}
{Foster}, A.~R., {Ji}, L., {Smith}, R.~K., \& {Brickhouse}, N.~S. 2012, \apj, 756, 128, \dodoi{10.1088/0004-637X/756/2/128}

\bibitem[{{Fulton} \& {Petigura}(2018)}]{fulton_2018}
{Fulton}, B.~J., \& {Petigura}, E.~A. 2018, \aj, 156, 264, \dodoi{10.3847/1538-3881/aae828}

\bibitem[{{Fulton} {et~al.}(2017){Fulton}, {Petigura}, {Howard}, {Isaacson}, {Marcy}, {Cargile}, {Hebb}, {Weiss}, {Johnson}, {Morton}, {Sinukoff}, {Crossfield}, \& {Hirsch}}]{fulton_2017}
{Fulton}, B.~J., {Petigura}, E.~A., {Howard}, A.~W., {et~al.} 2017, \aj, 154, 109, \dodoi{10.3847/1538-3881/aa80eb}

\bibitem[{{Gabriel} {et~al.}(2004){Gabriel}, {Denby}, {Fyfe}, {Hoar}, {Ibarra}, {Ojero}, {Osborne}, {Saxton}, {Lammers}, \& {Vacanti}}]{gabriel_2004}
{Gabriel}, C., {Denby}, M., {Fyfe}, D.~J., {et~al.} 2004, in Astronomical Society of the Pacific Conference Series, Vol. 314, Astronomical Data Analysis Software and Systems (ADASS) XIII, ed. F.~{Ochsenbein}, M.~G. {Allen}, \& D.~{Egret}, 759

\bibitem[{{Gao} {et~al.}(2023){Gao}, {Piette}, {Steinrueck}, {Nixon}, {Zhang}, {Kempton}, {Bean}, {Rauscher}, {Parmentier}, {Batalha}, {Savel}, {Arnold}, {Roman}, {Malsky}, \& {Taylor}}]{gao_2023}
{Gao}, P., {Piette}, A. A.~A., {Steinrueck}, M.~E., {et~al.} 2023, \apj, 951, 96, \dodoi{10.3847/1538-4357/acd16f}

\bibitem[{{Ginzburg} {et~al.}(2018){Ginzburg}, {Schlichting}, \& {Sari}}]{ginzburg_2018}
{Ginzburg}, S., {Schlichting}, H.~E., \& {Sari}, R. 2018, \mnras, 476, 759, \dodoi{10.1093/mnras/sty290}

\bibitem[{{Gupta} \& {Schlichting}(2019)}]{gupta_2019}
{Gupta}, A., \& {Schlichting}, H.~E. 2019, \mnras, 487, 24, \dodoi{10.1093/mnras/stz1230}

\bibitem[{{Hawthorn} {et~al.}(2023){Hawthorn}, {Bayliss}, {Wilson}, {Bonfanti}, {Adibekyan}, {Alibert}, {Sousa}, {Collins}, {Bryant}, {Osborn}, {Armstrong}, {Abe}, {Acton}, {Addison}, {Agabi}, {Alonso}, {Alves}, {Anglada-Escud{\'e}}, {B{\'a}rczy}, {Barclay}, {Barrado}, {Barros}, {Baumjohann}, {Bendjoya}, {Benz}, {Bieryla}, {Bonfils}, {Bouchy}, {Brandeker}, {Broeg}, {Brown}, {Burleigh}, {Buttu}, {Cabrera}, {Caldwell}, {Casewell}, {Charbonneau}, {Charnoz}, {Cloutier}, {Collier Cameron}, {Collins}, {Conti}, {Crouzet}, {Czismadia}, {Davies}, {Deleuil}, {Delgado-Mena}, {Delrez}, {Demangeon}, {Demory}, {Dransfield}, {Dumusque}, {Egger}, {Ehrenreich}, {Eigm{\"u}ller}, {Erickson}, {Essack}, {Fortier}, {Fossati}, {Fridlund}, {G{\"u}nther}, {G{\"u}del}, {Gandolfi}, {Gillard}, {Gillon}, {Gnilka}, {Goad}, {Goeke}, {Guillot}, {Hadjigeorghiou}, {Hellier}, {Henderson}, {Heng}, {Hooton}, {Horne}, {Howell}, {Hoyer}, {Irwin}, {Jenkins}, {Jenkins}, {Jensen}, {Kane}, {Kendall}, {Kielkopf}, {Kiss}, {Lacedelli}, {Laskar},
  {Latham}, {Etangs}, {Leleu}, {Lendl}, {Lillo-Box}, {Lovis}, {M{\'e}karnia}, {Massey}, {Masters}, {Maxted}, {Nascimbeni}, {Nielsen}, {O'Brien}, {Olofsson}, {Osborn}, {Pagano}, {Pall{\'e}}, {Persson}, {Piotto}, {Plavchan}, {Pollacco}, {Queloz}, {Ragazzoni}, {Rauer}, {Ribas}, {Ricker}, {S{\'e}gransan}, {Salmon}, {Santerne}, {Santos}, {Scandariato}, {Schmider}, {Schwarz}, {Seager}, {Shporer}, {Simon}, {Smith}, {Srdoc}, {Steller}, {Suarez}, {Szab{\'o}}, {Teske}, {Thomas}, {Tilbrook}, {Triaud}, {Udry}, {Van Grootel}, {Walton}, {Wang}, {Wheatley}, {Winn}, {Wittenmyer}, \& {Zhang}}]{hawthorn_2023}
{Hawthorn}, F., {Bayliss}, D., {Wilson}, T.~G., {et~al.} 2023, \mnras, 520, 3649, \dodoi{10.1093/mnras/stad306}

\bibitem[{Hu {et~al.}(2015)Hu, Seager, \& Yung}]{hu_2015}
Hu, R., Seager, S., \& Yung, Y.~L. 2015, ApJ, 807, 8, \dodoi{10.1088/0004-637X/807/1/8}

\bibitem[{{Hunter}(2007)}]{hunter_2007}
{Hunter}, J.~D. 2007, Computing in Science and Engineering, 9, 90, \dodoi{10.1109/MCSE.2007.55}

\bibitem[{{Husser} {et~al.}(2013){Husser}, {Wende-von Berg}, {Dreizler}, {Homeier}, {Reiners}, {Barman}, \& {Hauschildt}}]{husser_2013}
{Husser}, T.~O., {Wende-von Berg}, S., {Dreizler}, S., {et~al.} 2013, \aap, 553, A6, \dodoi{10.1051/0004-6361/201219058}

\bibitem[{Jackson {et~al.}(2012)Jackson, Davis, \& Wheatley}]{jackson_2012}
Jackson, A.~P., Davis, T.~A., \& Wheatley, P.~J. 2012, MNRAS, 422, 2024, \dodoi{10.1111/j.1365-2966.2012.20657.x}

\bibitem[{{Jensen}(2013)}]{Jensen:2013}
{Jensen}, E. 2013, {Tapir: A web interface for transit/eclipse observability}, Astrophysics Source Code Library.
\newblock \doeprint{1306.007}

\bibitem[{{Johnstone} {et~al.}(2021){Johnstone}, {Bartel}, \& {G{\"u}del}}]{johnstone_2021}
{Johnstone}, C.~P., {Bartel}, M., \& {G{\"u}del}, M. 2021, \aap, 649, A96, \dodoi{10.1051/0004-6361/202038407}

\bibitem[{{Johnstone} {et~al.}(2018){Johnstone}, {G{\"u}del}, {Lammer}, \& {Kislyakova}}]{johnstone_2018}
{Johnstone}, C.~P., {G{\"u}del}, M., {Lammer}, H., \& {Kislyakova}, K.~G. 2018, \aap, 617, A107, \dodoi{10.1051/0004-6361/201832776}

\bibitem[{Kempton {et~al.}(2023)Kempton, Zhang, Bean, Steinrueck, Piette, Parmentier, Malsky, Roman, Rauscher, Gao, Bell, Xue, Taylor, Savel, Arnold, Nixon, Stevenson, Mansfield, Kendrew, Zieba, Ducrot, Dyrek, Lagage, Stassun, Henry, Barman, Lupu, Malik, Kataria, Ih, Fu, Welbanks, \& McGill}]{kempton_2023}
Kempton, E. M.-R., Zhang, M., Bean, J.~L., {et~al.} 2023, Nature, \dodoi{10.1038/s41586-023-06159-5}

\bibitem[{Khodachenko {et~al.}(2015)Khodachenko, Shaikhislamov, Lammer, \& Prokopov}]{khodachenko_2015}
Khodachenko, M.~L., Shaikhislamov, I.~F., Lammer, H., \& Prokopov, P.~A. 2015, The Astrophysical Journal, 813, 50, \dodoi{10.1088/0004-637X/813/1/50}

\bibitem[{King \& Wheatley(2020)}]{king_2020}
King, G.~W., \& Wheatley, P.~J. 2020, Monthly Notices of the Royal Astronomical Society: Letters, 501, L28, \dodoi{10.1093/mnrasl/slaa186}

\bibitem[{{Kreidberg}(2015)}]{kreidberg_2015}
{Kreidberg}, L. 2015, \pasp, 127, 1161, \dodoi{10.1086/683602}

\bibitem[{{Lamp{\'o}n} {et~al.}(2020){Lamp{\'o}n}, {L{\'o}pez-Puertas}, {Lara}, {S{\'a}nchez-L{\'o}pez}, {Salz}, {Czesla}, {Sanz-Forcada}, {Molaverdikhani}, {Alonso-Floriano}, {Nortmann}, {Caballero}, {Bauer}, {Pall{\'e}}, {Montes}, {Quirrenbach}, {Nagel}, {Ribas}, {Reiners}, \& {Amado}}]{lampon_2020}
{Lamp{\'o}n}, M., {L{\'o}pez-Puertas}, M., {Lara}, L.~M., {et~al.} 2020, \aap, 636, A13, \dodoi{10.1051/0004-6361/201937175}

\bibitem[{{Lamp{\'o}n} {et~al.}(2021{\natexlab{a}}){Lamp{\'o}n}, {L{\'o}pez-Puertas}, {Sanz-Forcada}, {S{\'a}nchez-L{\'o}pez}, {Molaverdikhani}, {Czesla}, {Quirrenbach}, {Pall{\'e}}, {Caballero}, {Henning}, {Salz}, {Nortmann}, {Aceituno}, {Amado}, {Bauer}, {Montes}, {Nagel}, {Reiners}, \& {Ribas}}]{lampon_2021b}
{Lamp{\'o}n}, M., {L{\'o}pez-Puertas}, M., {Sanz-Forcada}, J., {et~al.} 2021{\natexlab{a}}, \aap, 647, A129, \dodoi{10.1051/0004-6361/202039417}

\bibitem[{{Lamp{\'o}n} {et~al.}(2021{\natexlab{b}}){Lamp{\'o}n}, {L{\'o}pez-Puertas}, {Czesla}, {S{\'a}nchez-L{\'o}pez}, {Lara}, {Salz}, {Sanz-Forcada}, {Molaverdikhani}, {Quirrenbach}, {Pall{\'e}}, {Caballero}, {Henning}, {Nortmann}, {Amado}, {Montes}, {Reiners}, \& {Ribas}}]{lampon_2021a}
{Lamp{\'o}n}, M., {L{\'o}pez-Puertas}, M., {Czesla}, S., {et~al.} 2021{\natexlab{b}}, \aap, 648, L7, \dodoi{10.1051/0004-6361/202140423}

\bibitem[{{Linssen} {et~al.}(2024){Linssen}, {Shih}, {MacLeod}, \& {Oklop{\v{c}}i{\'c}}}]{linssen_2024}
{Linssen}, D., {Shih}, J., {MacLeod}, M., \& {Oklop{\v{c}}i{\'c}}, A. 2024, \aap, 688, A43, \dodoi{10.1051/0004-6361/202450240}

\bibitem[{{Lithwick} {et~al.}(2012){Lithwick}, {Xie}, \& {Wu}}]{lithwick_2012}
{Lithwick}, Y., {Xie}, J., \& {Wu}, Y. 2012, \apj, 761, 122, \dodoi{10.1088/0004-637X/761/2/122}

\bibitem[{Luque \& Pallé(2022)}]{luque_2022}
Luque, R., \& Pallé, E. 2022, Science, 377, 1211, \dodoi{10.1126/science.abl7164}

\bibitem[{MacLeod \& Oklopčić(2022)}]{macleod_2022}
MacLeod, M., \& Oklopčić, A. 2022, The Astrophysical Journal, 926, 226, \dodoi{10.3847/1538-4357/ac46ce}

\bibitem[{{Madhusudhan} {et~al.}(2023){Madhusudhan}, {Sarkar}, {Constantinou}, {Holmberg}, {Piette}, \& {Moses}}]{madhusudhan_2023}
{Madhusudhan}, N., {Sarkar}, S., {Constantinou}, S., {et~al.} 2023, \apjl, 956, L13, \dodoi{10.3847/2041-8213/acf577}

\bibitem[{{Magic} {et~al.}(2013){Magic}, {Collet}, {Asplund}, {Trampedach}, {Hayek}, {Chiavassa}, {Stein}, \& {Nordlund}}]{magic_2013}
{Magic}, Z., {Collet}, R., {Asplund}, M., {et~al.} 2013, \aap, 557, A26, \dodoi{10.1051/0004-6361/201321274}

\bibitem[{{McCully} {et~al.}(2018){McCully}, {Volgenau}, {Harbeck}, {Lister}, {Saunders}, {Turner}, {Siiverd}, \& {Bowman}}]{McCully:2018}
{McCully}, C., {Volgenau}, N.~H., {Harbeck}, D.-R., {et~al.} 2018, in Society of Photo-Optical Instrumentation Engineers (SPIE) Conference Series, Vol. 10707, \procspie, 107070K, \dodoi{10.1117/12.2314340}

\bibitem[{{Merloni} {et~al.}(2024){Merloni}, {Lamer}, {Liu}, {Ramos-Ceja}, {Brunner}, {Bulbul}, {Dennerl}, {Doroshenko}, {Freyberg}, {Friedrich}, {Gatuzz}, {Georgakakis}, {Haberl}, {Igo}, {Kreykenbohm}, {Liu}, {Maitra}, {Malyali}, {Mayer}, {Nandra}, {Predehl}, {Robrade}, {Salvato}, {Sanders}, {Stewart}, {Tub{\'\i}n-Arenas}, {Weber}, {Wilms}, {Arcodia}, {Artis}, {Aschersleben}, {Avakyan}, {Aydar}, {Bahar}, {Balzer}, {Becker}, {Berger}, {Boller}, {Bornemann}, {Br{\"u}ggen}, {Brusa}, {Buchner}, {Burwitz}, {Camilloni}, {Clerc}, {Comparat}, {Coutinho}, {Czesla}, {Dannhauer}, {Dauner}, {Dauser}, {Dietl}, {Dolag}, {Dwelly}, {Egg}, {Ehl}, {Freund}, {Friedrich}, {Gaida}, {Garrel}, {Ghirardini}, {Gokus}, {Gr{\"u}nwald}, {Grandis}, {Grotova}, {Gruen}, {Gueguen}, {H{\"a}mmerich}, {Hamaus}, {Hasinger}, {Haubner}, {Homan}, {Ider Chitham}, {Joseph}, {Joyce}, {K{\"o}nig}, {Kaltenbrunner}, {Khokhriakova}, {Kink}, {Kirsch}, {Kluge}, {Knies}, {Krippendorf}, {Krumpe}, {Kurpas}, {Li}, {Liu}, {Locatelli}, {Lorenz}, {M{\"u}ller},
  {Magaudda}, {Mannes}, {McCall}, {Meidinger}, {Michailidis}, {Migkas}, {Mu{\~n}oz-Giraldo}, {Musiimenta}, {Nguyen-Dang}, {Ni}, {Olechowska}, {Ota}, {Pacaud}, {Pasini}, {Perinati}, {Pires}, {Pommranz}, {Ponti}, {Poppenhaeger}, {P{\"u}hlhofer}, {Rau}, {Reh}, {Reiprich}, {Roster}, {Saeedi}, {Santangelo}, {Sasaki}, {Schmitt}, {Schneider}, {Schrabback}, {Schuster}, {Schwope}, {Seppi}, {Serim}, {Shreeram}, {Sokolova-Lapa}, {Starck}, {Stelzer}, {Stierhof}, {Suleimanov}, {Tenzer}, {Traulsen}, {Tr{\"u}mper}, {Tsuge}, {Urrutia}, {Veronica}, {Waddell}, {Willer}, {Wolf}, {Yeung}, {Zainab}, {Zangrandi}, {Zhang}, {Zhang}, \& {Zheng}}]{merloni_2024}
{Merloni}, A., {Lamer}, G., {Liu}, T., {et~al.} 2024, \aap, 682, A34, \dodoi{10.1051/0004-6361/202347165}

\bibitem[{{Mignone} {et~al.}(2007){Mignone}, {Bodo}, {Massaglia}, {Matsakos}, {Tesileanu}, {Zanni}, \& {Ferrari}}]{mignone_2007}
{Mignone}, A., {Bodo}, G., {Massaglia}, S., {et~al.} 2007, \apjs, 170, 228, \dodoi{10.1086/513316}

\bibitem[{{Mills} \& {Mazeh}(2017)}]{mills_2017}
{Mills}, S.~M., \& {Mazeh}, T. 2017, \apjl, 839, L8, \dodoi{10.3847/2041-8213/aa67eb}

\bibitem[{{Modirrousta-Galian} \& {Korenaga}(2024)}]{modirrousta-galian_2024}
{Modirrousta-Galian}, D., \& {Korenaga}, J. 2024, \apj, 965, 97, \dodoi{10.3847/1538-4357/ad276f}

\bibitem[{Mousis {et~al.}(2020)Mousis, Deleuil, Aguichine, Marcq, Naar, Aguirre, Brugger, \& Gonçalves}]{mousis_2020}
Mousis, O., Deleuil, M., Aguichine, A., {et~al.} 2020, ApJL, 896, L22, \dodoi{10.3847/2041-8213/ab9530}

\bibitem[{{Murray-Clay} {et~al.}(2009){Murray-Clay}, {Chiang}, \& {Murray}}]{murray-clay_2009}
{Murray-Clay}, R.~A., {Chiang}, E.~I., \& {Murray}, N. 2009, \apj, 693, 23, \dodoi{10.1088/0004-637X/693/1/23}

\bibitem[{{Oklop{\v{c}}i{\'c}}(2019)}]{oklopcic_2019}
{Oklop{\v{c}}i{\'c}}, A. 2019, \apj, 881, 133, \dodoi{10.3847/1538-4357/ab2f7f}

\bibitem[{Oklopčić \& Hirata(2018)}]{oklopcic_2018}
Oklopčić, A., \& Hirata, C.~M. 2018, ApJL, 855, L11, \dodoi{10.3847/2041-8213/aaada9}

\bibitem[{{Orell-Miquel} {et~al.}(2023){Orell-Miquel}, {Lamp{\'o}n}, {L{\'o}pez-Puertas}, {Mallorqu{\'\i}n}, {Murgas}, {Pel{\'a}ez-Torres}, {Pall{\'e}}, {Esparza-Borges}, {Sanz-Forcada}, {Tabernero}, {Nortmann}, {Nagel}, {Parviainen}, {Zapatero Osorio}, {Caballero}, {Czesla}, {Cifuentes}, {Morello}, {Quirrenbach}, {Amado}, {Fern{\'a}ndez-Mart{\'\i}n}, {Fukui}, {Henning}, {Kawauchi}, {de Leon}, {Molaverdikhani}, {Montes}, {Narita}, {Reiners}, {Ribas}, {S{\'a}nchez-L{\'o}pez}, {Schweitzer}, {Stangret}, \& {Yan}}]{orell-miquel_2023}
{Orell-Miquel}, J., {Lamp{\'o}n}, M., {L{\'o}pez-Puertas}, M., {et~al.} 2023, \aap, 677, A56, \dodoi{10.1051/0004-6361/202346445}

\bibitem[{{Orell-Miquel} {et~al.}(2024){Orell-Miquel}, {Murgas}, {Pall{\'e}}, {Mallorqu{\'\i}n}, {L{\'o}pez-Puertas}, {Lamp{\'o}n}, {Sanz-Forcada}, {Nortmann}, {Czesla}, {Nagel}, {Ribas}, {Stangret}, {Livingston}, {Knudstrup}, {Albrecht}, {Carleo}, {Caballero}, {Dai}, {Esparza-Borges}, {Fukui}, {Heng}, {Henning}, {Kagetani}, {Lesjak}, {de Leon}, {Montes}, {Morello}, {Narita}, {Quirrenbach}, {Amado}, {Reiners}, {Schweitzer}, \& {Vico Linares}}]{orell-miquel_2024}
{Orell-Miquel}, J., {Murgas}, F., {Pall{\'e}}, E., {et~al.} 2024, \aap, 689, A179, \dodoi{10.1051/0004-6361/202449411}

\bibitem[{{Owen}(2019)}]{owen_2019}
{Owen}, J.~E. 2019, Annual Review of Earth and Planetary Sciences, 47, 67, \dodoi{10.1146/annurev-earth-053018-060246}

\bibitem[{{Owen} \& {Adams}(2014)}]{owen_2014}
{Owen}, J.~E., \& {Adams}, F.~C. 2014, \mnras, 444, 3761, \dodoi{10.1093/mnras/stu1684}

\bibitem[{{Owen} \& {Schlichting}(2024)}]{owen_2024}
{Owen}, J.~E., \& {Schlichting}, H.~E. 2024, \mnras, 528, 1615, \dodoi{10.1093/mnras/stad3972}

\bibitem[{{Owen} \& {Wu}(2017)}]{owen_2017}
{Owen}, J.~E., \& {Wu}, Y. 2017, \apj, 847, 29, \dodoi{10.3847/1538-4357/aa890a}

\bibitem[{{Owen} {et~al.}(2023){Owen}, {Murray-Clay}, {Schreyer}, {Schlichting}, {Ardila}, {Gupta}, {Loyd}, {Shkolnik}, {Sing}, \& {Swain}}]{owen_2023}
{Owen}, J.~E., {Murray-Clay}, R.~A., {Schreyer}, E., {et~al.} 2023, \mnras, 518, 4357, \dodoi{10.1093/mnras/stac3414}

\bibitem[{{Parker}(1958)}]{parker_1958}
{Parker}, E.~N. 1958, \apj, 128, 664, \dodoi{10.1086/146579}

\bibitem[{Rogers {et~al.}(2023)Rogers, Schlichting, \& Owen}]{rogers_2023}
Rogers, J.~G., Schlichting, H.~E., \& Owen, J.~E. 2023, The Astrophysical Journal Letters, 947, L19, \dodoi{10.3847/2041-8213/acc86f}

\bibitem[{{Rosener} {et~al.}(2025){Rosener}, {Zhang}, \& {Bean}}]{rosener_2025}
{Rosener}, R., {Zhang}, M., \& {Bean}, J.~L. 2025, \apj, 980, 34, \dodoi{10.3847/1538-4357/ada6ab}

\bibitem[{{Salz} {et~al.}(2015){Salz}, {Banerjee}, {Mignone}, {Schneider}, {Czesla}, \& {Schmitt}}]{salz_2015}
{Salz}, M., {Banerjee}, R., {Mignone}, A., {et~al.} 2015, \aap, 576, A21, \dodoi{10.1051/0004-6361/201424330}

\bibitem[{{Sanz-Forcada} {et~al.}(2025){Sanz-Forcada}, {L{\'o}pez-Puertas}, {Lamp{\'o}n}, {Czesla}, {Nortmann}, {Caballero}, {Zapatero Osorio}, {Amado}, {Murgas}, {Orell-Miquel}, {Pall{\'e}}, {Quirrenbach}, {Reiners}, {Ribas}, {S{\'a}nchez-L{\'o}pez}, \& {Solano}}]{sanz-forcada_2025}
{Sanz-Forcada}, J., {L{\'o}pez-Puertas}, M., {Lamp{\'o}n}, M., {et~al.} 2025, \aap, 693, A285, \dodoi{10.1051/0004-6361/202451680}

\bibitem[{{Schreyer} {et~al.}(2024){Schreyer}, {Owen}, {Loyd}, \& {Murray-Clay}}]{schreyer_2024}
{Schreyer}, E., {Owen}, J.~E., {Loyd}, R.~O.~P., \& {Murray-Clay}, R. 2024, \mnras, 533, 3296, \dodoi{10.1093/mnras/stae1976}

\bibitem[{{Schulik} \& {Owen}(2024)}]{schulik_2024}
{Schulik}, M., \& {Owen}, J. 2024, arXiv e-prints, arXiv:2412.05258, \dodoi{10.48550/arXiv.2412.05258}

\bibitem[{{Smette} {et~al.}(2015){Smette}, {Sana}, {Noll}, {Horst}, {Kausch}, {Kimeswenger}, {Barden}, {Szyszka}, {Jones}, {Gallenne}, {Vinther}, {Ballester}, \& {Taylor}}]{smette_2015}
{Smette}, A., {Sana}, H., {Noll}, S., {et~al.} 2015, \aap, 576, A77, \dodoi{10.1051/0004-6361/201423932}

\bibitem[{{Smith} {et~al.}(2001){Smith}, {Brickhouse}, {Liedahl}, \& {Raymond}}]{smith_2001}
{Smith}, R.~K., {Brickhouse}, N.~S., {Liedahl}, D.~A., \& {Raymond}, J.~C. 2001, \apj, 556, L91, \dodoi{10.1086/322992}

\bibitem[{{Spake} {et~al.}(2018){Spake}, {Sing}, {Evans}, {Oklop{\v{c}}i{\'c}}, {Bourrier}, {Kreidberg}, {Rackham}, {Irwin}, {Ehrenreich}, {Wyttenbach}, {Wakeford}, {Zhou}, {Chubb}, {Nikolov}, {Goyal}, {Henry}, {Williamson}, {Blumenthal}, {Anderson}, {Hellier}, {Charbonneau}, {Udry}, \& {Madhusudhan}}]{spake_2018}
{Spake}, J.~J., {Sing}, D.~K., {Evans}, T.~M., {et~al.} 2018, \nat, 557, 68, \dodoi{10.1038/s41586-018-0067-5}

\bibitem[{Speagle(2020)}]{speagle_2019}
Speagle, J.~S. 2020, MNRAS, 493, 3132, \dodoi{10.1093/mnras/staa278}

\bibitem[{{van der Walt} {et~al.}(2011){van der Walt}, {Colbert}, \& {Varoquaux}}]{van_der_walt_2011}
{van der Walt}, S., {Colbert}, S.~C., \& {Varoquaux}, G. 2011, Computing in Science and Engineering, 13, 22, \dodoi{10.1109/MCSE.2011.37}

\bibitem[{{Vidotto} {et~al.}(2023){Vidotto}, {Bourrier}, {Fares}, {Bellotti}, {Donati}, {Petit}, {Hussain}, \& {Morin}}]{vidotto_2023}
{Vidotto}, A.~A., {Bourrier}, V., {Fares}, R., {et~al.} 2023, \aap, 678, A152, \dodoi{10.1051/0004-6361/202347237}

\bibitem[{{Virtanen} {et~al.}(2020){Virtanen}, {Gommers}, {Oliphant}, {Haberland}, {Reddy}, {Cournapeau}, {Burovski}, {Peterson}, {Weckesser}, {Bright}, {van der Walt}, {Brett}, {Wilson}, {Jarrod Millman}, {Mayorov}, {Nelson}, {Jones}, {Kern}, {Larson}, {Carey}, {Polat}, {Feng}, {Moore}, {Vand erPlas}, {Laxalde}, {Perktold}, {Cimrman}, {Henriksen}, {Quintero}, {Harris}, {Archibald}, {Ribeiro}, {Pedregosa}, {van Mulbregt}, \& {Contributors}}]{virtanen_2020}
{Virtanen}, P., {Gommers}, R., {Oliphant}, T.~E., {et~al.} 2020, Nature Methods, 17, 261, \dodoi{https://doi.org/10.1038/s41592-019-0686-2}

\bibitem[{{Wallack} {et~al.}(2024){Wallack}, {Batalha}, {Alderson}, {Scarsdale}, {Adams Redai}, {Aguichine}, {Alam}, {Gao}, {Wolfgang}, {Batalha}, {Kirk}, {L{\'o}pez-Morales}, {Moran}, {Teske}, {Wakeford}, \& {Wogan}}]{wallack_2024}
{Wallack}, N.~L., {Batalha}, N.~E., {Alderson}, L., {et~al.} 2024, \aj, 168, 77, \dodoi{10.3847/1538-3881/ad3917}

\bibitem[{{Winn}(2010)}]{winn_2010}
{Winn}, J.~N. 2010, in Exoplanets, ed. S.~{Seager}, 55--77, \dodoi{10.48550/arXiv.1001.2010}

\bibitem[{{Wogan} {et~al.}(2024){Wogan}, {Batalha}, {Zahnle}, {Krissansen-Totton}, {Tsai}, \& {Hu}}]{wogan_2024}
{Wogan}, N.~F., {Batalha}, N.~E., {Zahnle}, K.~J., {et~al.} 2024, \apjl, 963, L7, \dodoi{10.3847/2041-8213/ad2616}

\bibitem[{{Wood} {et~al.}(2021){Wood}, {M{\"u}ller}, {Redfield}, {Konow}, {Vannier}, {Linsky}, {Youngblood}, {Vidotto}, {Jardine}, {Alvarado-G{\'o}mez}, \& {Drake}}]{wood_2021}
{Wood}, B.~E., {M{\"u}ller}, H.-R., {Redfield}, S., {et~al.} 2021, \apj, 915, 37, \dodoi{10.3847/1538-4357/abfda5}

\bibitem[{{Wordsworth} \& {Kreidberg}(2022)}]{wordsworth_2022}
{Wordsworth}, R., \& {Kreidberg}, L. 2022, \araa, 60, 159, \dodoi{10.1146/annurev-astro-052920-125632}

\bibitem[{Wright {et~al.}(2011)Wright, Drake, Mamajek, \& Henry}]{wright_2011}
Wright, N.~J., Drake, J.~J., Mamajek, E.~E., \& Henry, G.~W. 2011, ApJ, 743, 48, \dodoi{10.1088/0004-637X/743/1/48}

\bibitem[{Zahnle \& Kasting(1986)}]{zahnle_1986}
Zahnle, K.~J., \& Kasting, J.~F. 1986, Icarus, 68, 462, \dodoi{https://doi.org/10.1016/0019-1035(86)90051-5}

\bibitem[{{Zhang} {et~al.}(2023{\natexlab{a}}){Zhang}, {Dai}, {Bean}, {Knutson}, \& {Rescigno}}]{zhang_2023b}
{Zhang}, M., {Dai}, F., {Bean}, J.~L., {Knutson}, H.~A., \& {Rescigno}, F. 2023{\natexlab{a}}, \apjl, 953, L25, \dodoi{10.3847/2041-8213/aced51}

\bibitem[{{Zhang} {et~al.}(2023{\natexlab{b}}){Zhang}, {Knutson}, {Dai}, {Wang}, {Ricker}, {Schwarz}, {Mann}, \& {Collins}}]{zhang_2023}
{Zhang}, M., {Knutson}, H.~A., {Dai}, F., {et~al.} 2023{\natexlab{b}}, \aj, 165, 62, \dodoi{10.3847/1538-3881/aca75b}

\bibitem[{Zhang {et~al.}(2022)Zhang, Knutson, Wang, Dai, \& Barragán}]{zhang_2022b}
Zhang, M., Knutson, H.~A., Wang, L., Dai, F., \& Barragán, O. 2022, AJ, 163, 67, \dodoi{10.3847/1538-3881/ac3fa7}

\bibitem[{{Zhang} {et~al.}(2021){Zhang}, {Knutson}, {Wang}, {Dai}, {Oklopcic}, \& {Hu}}]{zhang_2021}
{Zhang}, M., {Knutson}, H.~A., {Wang}, L., {et~al.} 2021, \aj, 161, 181, \dodoi{10.3847/1538-3881/abe382}

\bibitem[{Zhu \& Dong(2021)}]{zhu_2021}
Zhu, W., \& Dong, S. 2021, Annual Review of Astronomy and Astrophysics, 59, 291, \dodoi{https://doi.org/10.1146/annurev-astro-112420-020055}

\end{thebibliography}
\end{document}